%% file: main.tex
\documentclass[sigplan]{acmart}
\renewcommand\footnotetextcopyrightpermission[1]{} 
\makeatletter                   
\pdfoutput=1
\def\mdseries@tt{m}             
\makeatother                    
\usepackage[plain]{fancyref}
\usepackage[draft=true]{minted} 
\usepackage{color}
\usepackage{hyperref}           
\hypersetup{
    colorlinks=true,
    linkcolor=blue,
    filecolor=red,      
    urlcolor=magenta,
    breaklinks=true,            
}
\usepackage{breakurl}           

\newcommand{\ibmqpoughkeepsie} {{IBMQ Poughkeepsie}}
\newcommand{\ibmqsystemone} {{IBMQ Johannesburg}}
\newcommand{\ibmqboeblingen} {{IBMQ Boeblingen}}

\newcommand{\numdevices} {{three}}
\newcommand{\seriessched} {{SerialSched}}
\newcommand{\parsched} {{ParSched}}
\newcommand{\xtalksched} {{XtalkSched}}
\newcommand{\hide}[1] {}

\usepackage[normalem]{ulem}
\usepackage{amssymb}
\usepackage{amsmath}
\usepackage{mathtools}
\usepackage[]{caption}

\usepackage{subfig}
\usepackage[]{algorithm2e}
\usepackage{multirow}
\usepackage{graphicx}
\usepackage{mathtools}
\usepackage{physics}
\usepackage{listings}
\usepackage{minted}
\usepackage{multicol}
\usepackage{booktabs}
\usepackage{xspace}
\usepackage{multicol}

\title{Software Mitigation of Crosstalk on \\ Noisy Intermediate-Scale Quantum Computers
}

\author{Prakash Murali \enskip David C. McKay* \enskip Margaret Martonosi \enskip Ali Javadi-Abhari*}
\affiliation{%
  \institution{Princeton University \enskip *IBM T. J. Watson Research Center}
}
\renewcommand{\authors}{Prakash Murali, David C. McKay, Margaret Martonosi, Ali Javadi-Abhari}

\date{}

\input{txt/abstract.tex}

\begin{document}
\maketitle
\input{txt/intro_new.tex}
\input{txt/background.tex}
\input{txt/related_work.tex}
\input{txt/designquestions.tex}
\input{txt/body.tex}
\input{txt/scheduler.tex}
\input{txt/expt.tex}

\input{txt/results.tex}
\input{txt/conclusions.tex}

\begin{acks}
This work is funded in part by EPiQC, an NSF Expedition in
Computing, under grants CCF-1730082. We thank 
Douglas T. McClure and Sarah Sheldon from IBM for useful discussions. We also thank the anonymous reviewers for their comments which helped improve the experiments.
\end{acks}
\bibliographystyle{ACM-Reference-Format}

\end{document}

%% file: txt/abstract.tex
\begin{abstract}
Crosstalk is a major source of noise in Noisy Intermediate-Scale Quantum (NISQ) systems and is a fundamental challenge for hardware design. When multiple instructions are executed in parallel, crosstalk between the instructions can corrupt the quantum state and lead to incorrect program execution. Our goal is to mitigate the application impact of crosstalk noise through software techniques. This requires (i) accurate characterization of hardware crosstalk, and (ii) intelligent instruction scheduling to serialize the affected operations. Since crosstalk characterization is computationally expensive, we develop optimizations which reduce the characterization overhead. On \numdevices\ 20-qubit IBMQ systems, we demonstrate two orders of magnitude reduction in characterization time (compute time on the QC device) compared to all-pairs crosstalk measurements. 
Informed by these characterization, we develop a scheduler that judiciously serializes high crosstalk instructions balancing the need to mitigate crosstalk and exponential decoherence errors from serialization. On real-system runs on \numdevices\ IBMQ systems, our scheduler improves the error rate of application circuits by up to 5.6x, compared to the IBM instruction scheduler and offers near-optimal crosstalk mitigation in practice.

In a broader picture, the difficulty of mitigating crosstalk has recently driven QC vendors to move towards sparser qubit connectivity or disabling nearby operations entirely in hardware, which can be detrimental to performance. Our work makes the case for software mitigation of crosstalk errors.

\end{abstract}

%% file: txt/intro_new.tex
\section{Introduction}
\begin{figure*}[t]
    \centering
    \subfloat[Machine]{
    \includegraphics[scale=0.3]{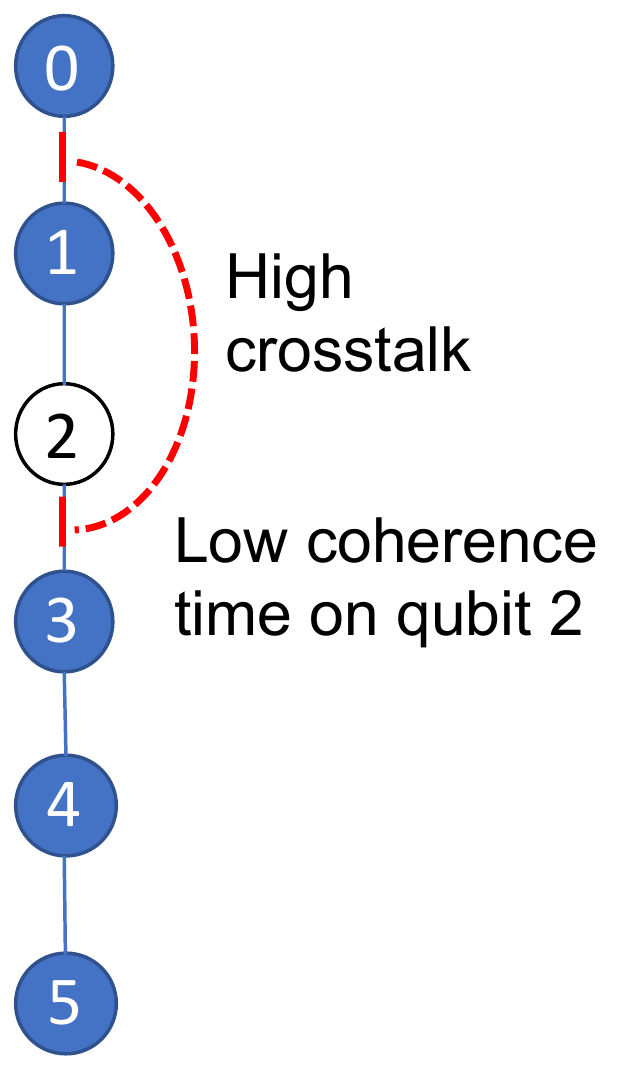}
    \label{fig:intro_example_machine}
    }
    \subfloat[Program IR]{
    \includegraphics[scale=0.2]{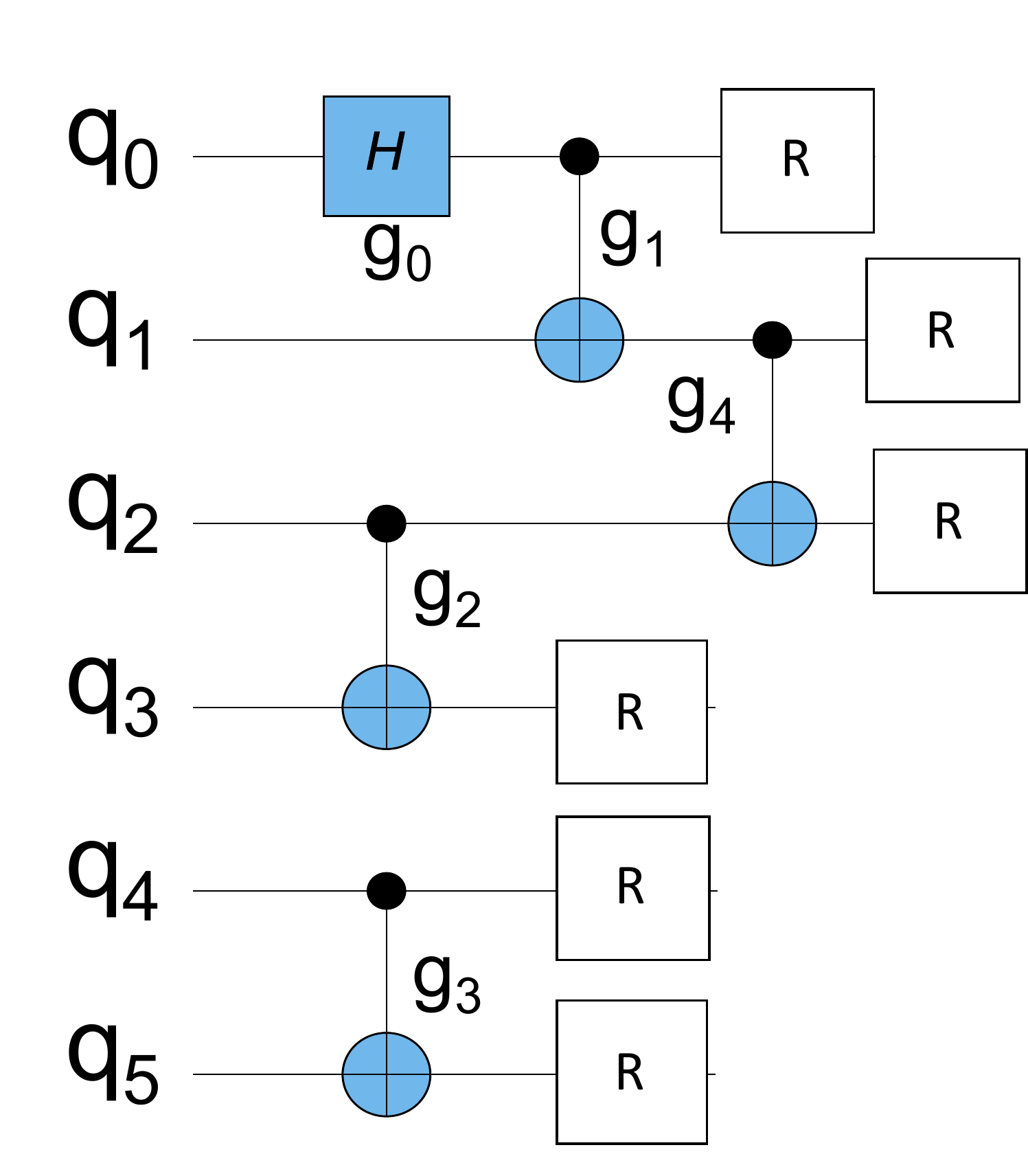}
    \label{fig:intro_example_prog}
    }
    \subfloat[Original Default Schedule]{
    \includegraphics[scale=0.2]{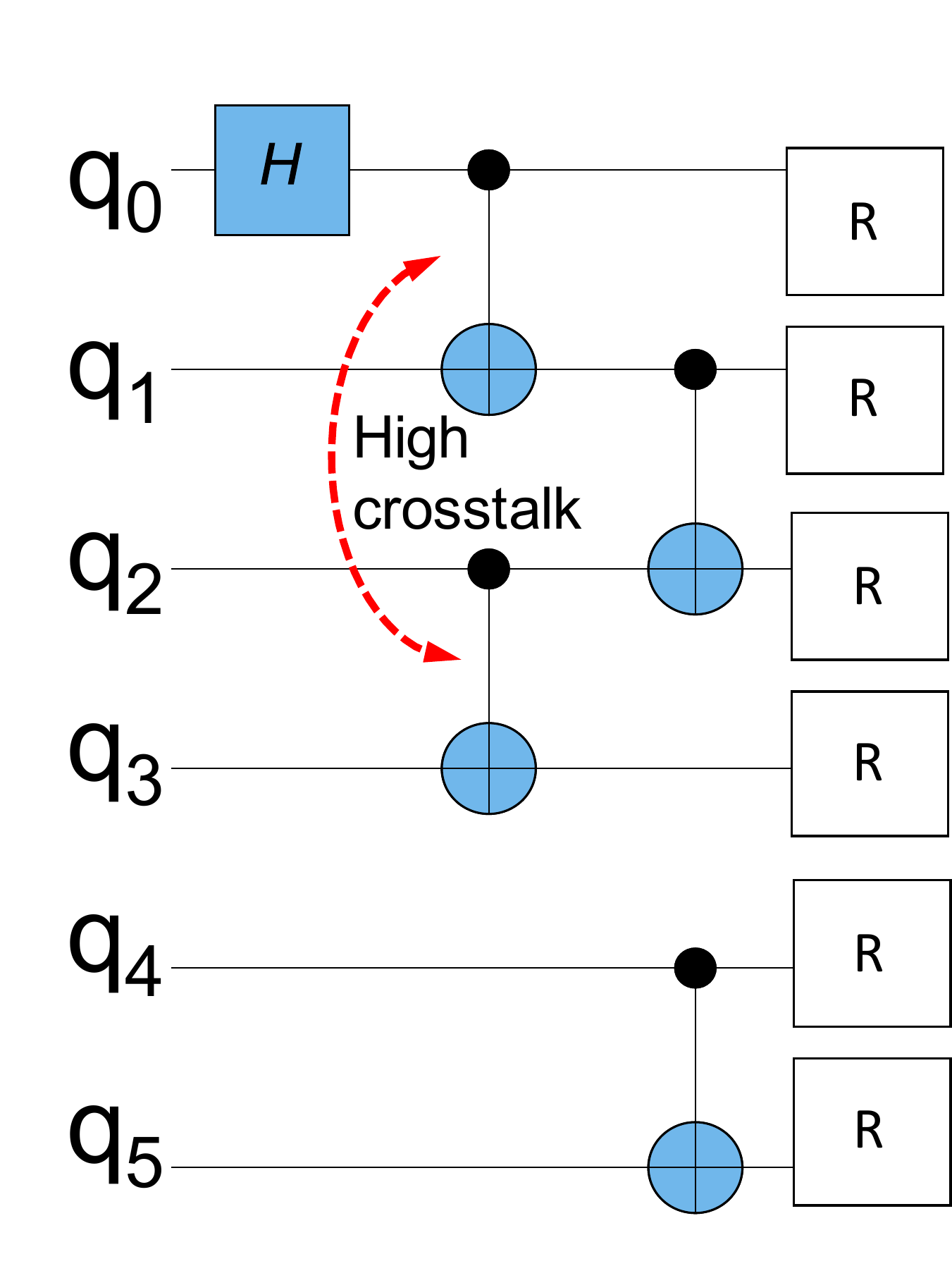}
    \label{figure:intro_ibm_schedule}
    }
    \subfloat[High decoherence schedule]{
    \includegraphics[scale=0.2]{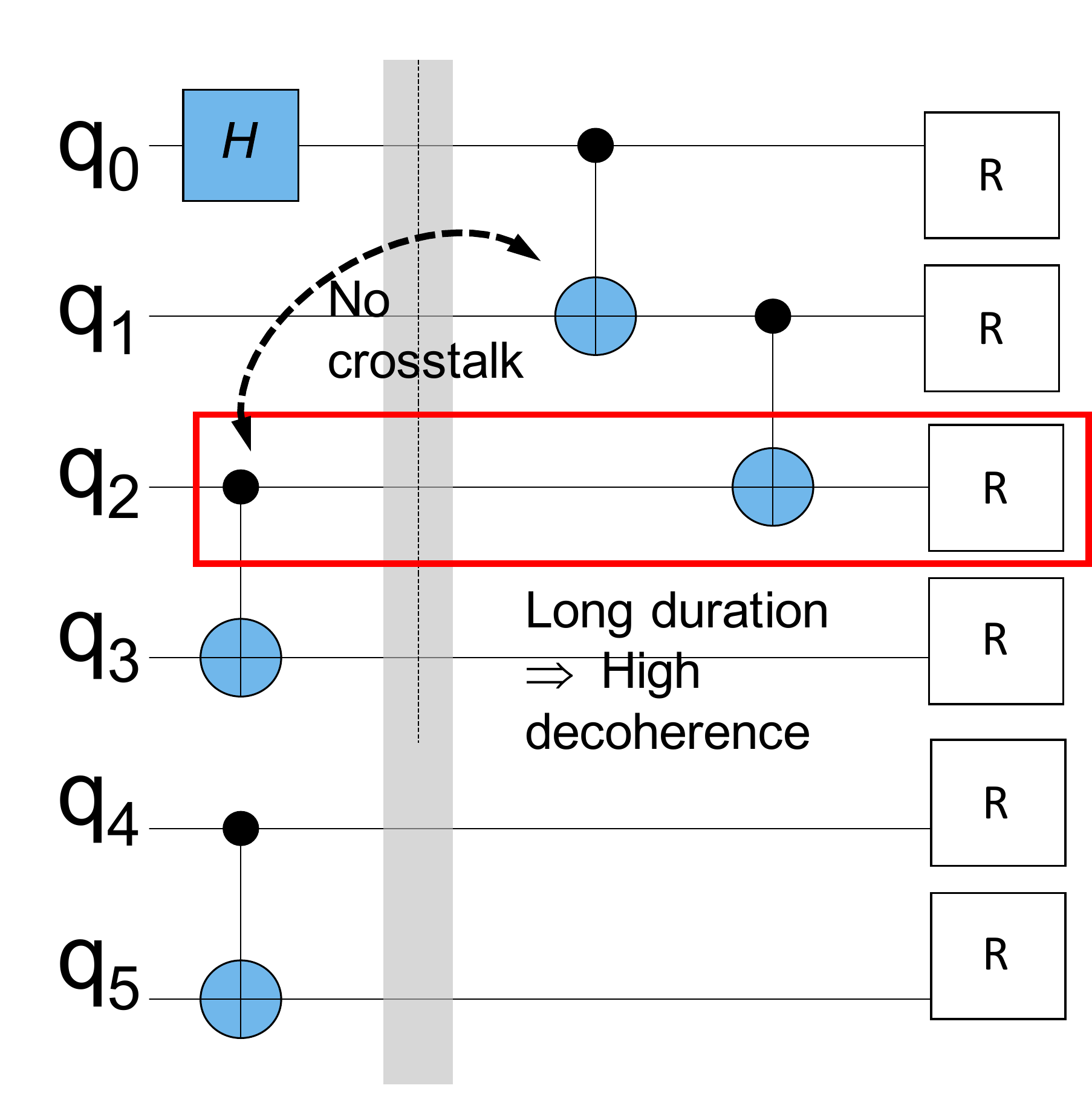}
    }
    \subfloat[Desired Schedule]{
    \includegraphics[scale=0.2]{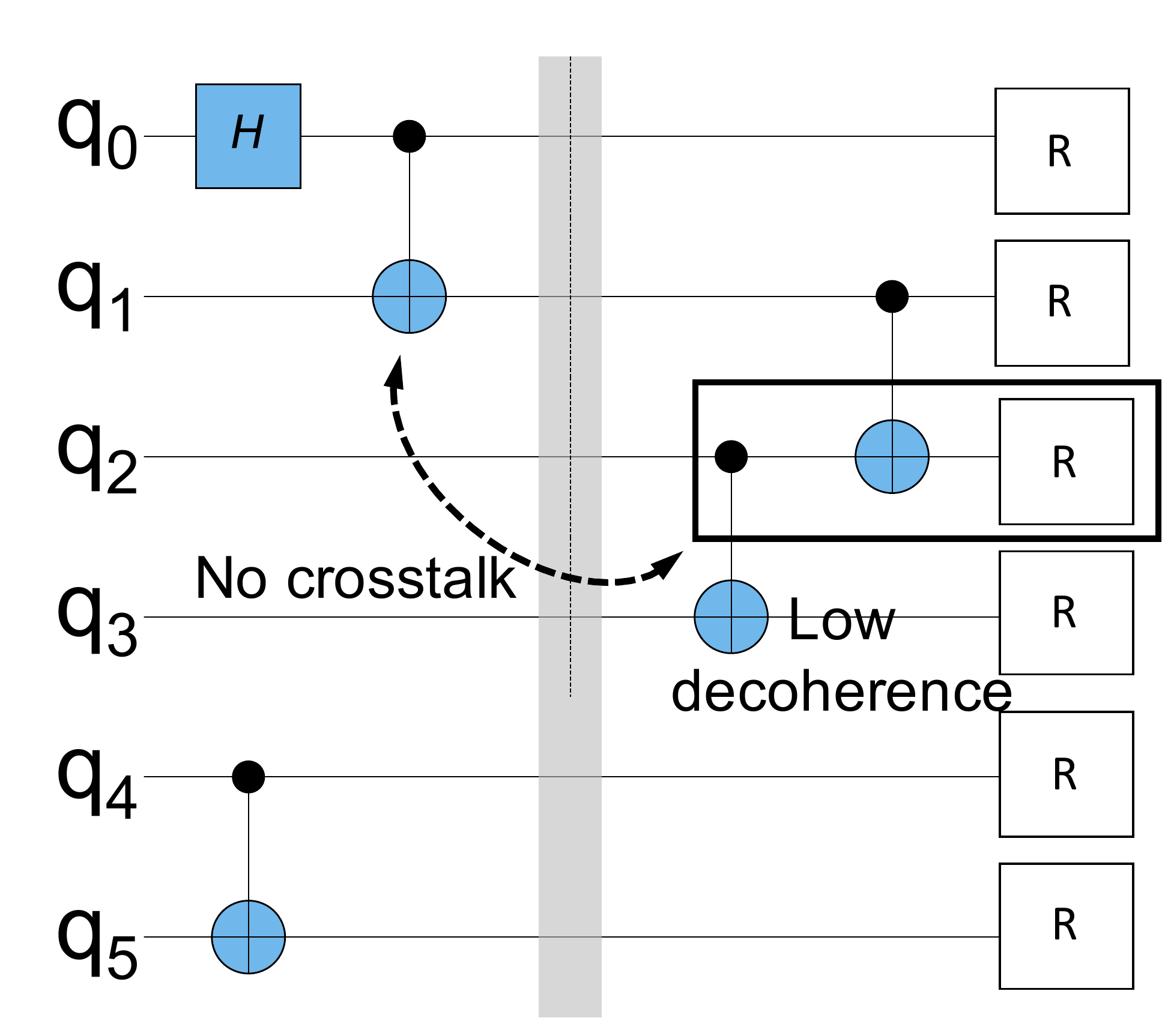}
    \label{fig:intro_example_desired}
    }
    \caption{(a) An example 6-qubit system. Nodes are physical qubits, and edges are possible CNOT gates. When a CNOT is executed on qubits (0, 1) and another CNOT is executed simultaneously on qubits (2, 3), the error rate of both CNOTs increases because of crosstalk.  Qubit 2 has low coherence, which means that long computation on that qubit (including any idle time after the first operation) is highly error prone. (b) An example program IR with parallelized operations. Dangling XOR operations are CNOTs and R is for readout. Time goes left to right. (c) Default schedule for this program on IBM hardware --- the schedule maximizes instruction parallelism, but the hardware is restricted to perform all readouts at the same time, so by default, all gates are right-aligned by the hardware scheduler. This schedule suffers from high crosstalk errors. (d) A schedule where the high crosstalk operations are naively serialized, leading to high decoherence error on qubit 2. (e) The desired schedule which avoids high crosstalk and high decoherence errors.}
    \label{fig:intro_example}
\end{figure*}
\begin{figure}
    \centering
`    \includegraphics[scale=0.32]{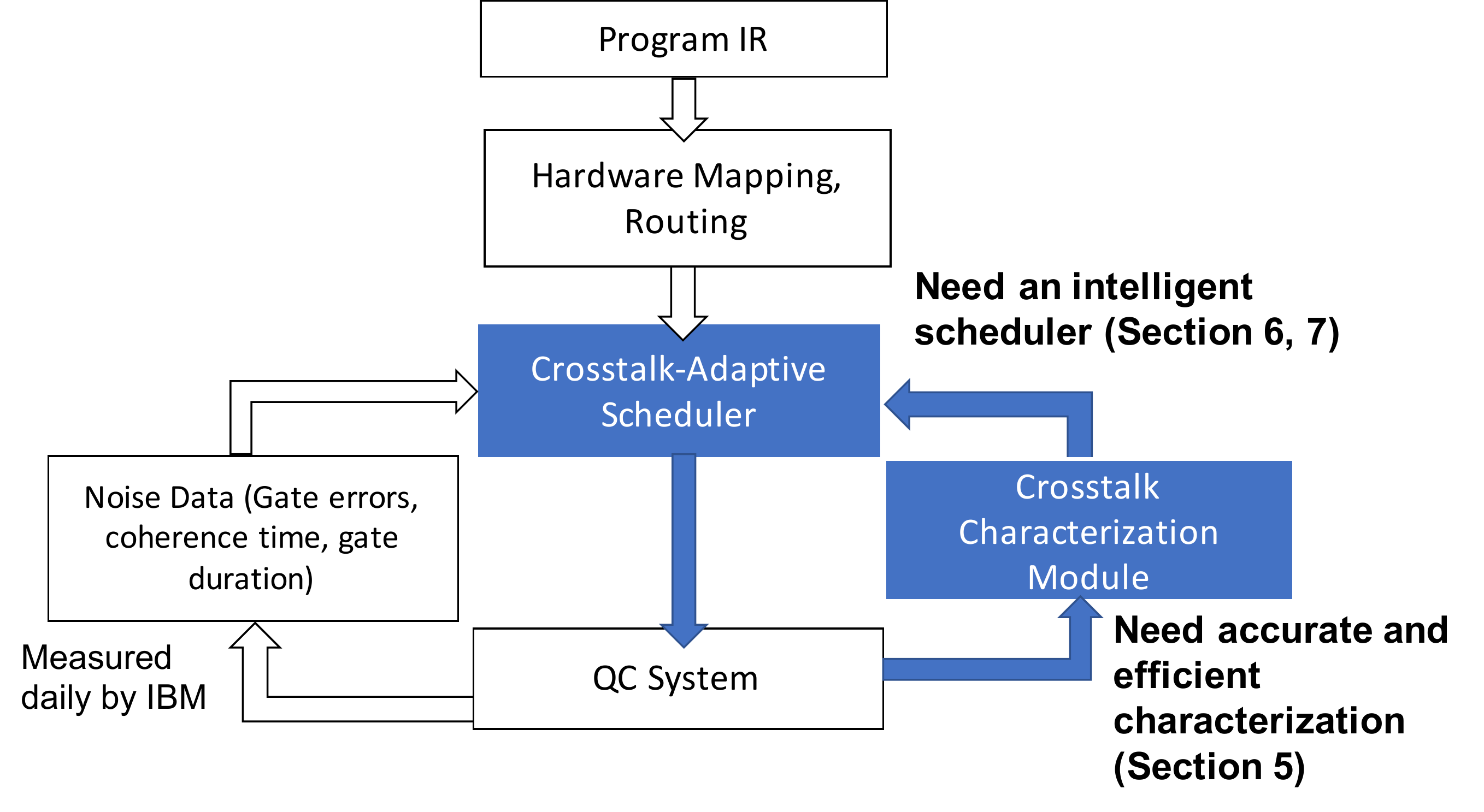}
    \caption{Our crosstalk mitigation approach. We develop two building blocks necessary for software mitigation of crosstalk. The first module performs fast and accurate characterization of the crosstalk noise present in the hardware. The second module performs instruction scheduling using the characterization data. Our scheduler serializes high crosstalk instructions but also balances the need to avoid exponential decoherence errors from serialization.}
    \label{fig:crosstalk_mitigation_flow}
\end{figure}
Quantum computing (QC) is a radically new paradigm of computing, where information is stored in \emph{qubits} (quantum bits) and manipulated using instructions known as \emph{gates}. By using quantum effects to efficiently navigate an exponentially-scaling state space, QC systems can arrive at a solution much faster for certain classically-intractable problems. 
This computational model has proven effective in diverse fields such as cryptography \cite{shor1}, chemistry \cite{vqe2, vqe1} and machine learning \cite{quantum_ml1, quantum_ml2}.

QC hardware has progressed considerably in the last few years. Prototype systems with 5-20 qubits are now available for broad public use~\cite{ibmq} and larger systems with 49-72 qubits are under development or test~\cite{googlebristlecone,ibm50q,intelq}. The term Noisy Intermediate-Scale Quantum (NISQ) refers to these near-term quantum systems with up to a few hundred qubits and imperfect qubits, gates and readout~\cite{nisq}. While too small and noisy to run large applications, they can support small application studies and are an important milestone on the way to practical QC. Compiler toolflows which optimize programs to make the best use of scarce hardware resources and mitigate the effects of hardware noise are therefore critical for useful computation on NISQ-era devices.

{\em Crosstalk} is a major source of noise in NISQ systems which corrupts the quantum state when multiple gates (instructions) are executed simultaneously. Crosstalk arises from fundamental challenges in QC hardware design such as unwanted interactions between the qubits and from leakage of the control signals (used to operate the gate) onto qubits which are not part of the intended gate operation. Crosstalk noise is prevalent across many of the leading qubits including superconducting and trapped ion qubits~\cite{xtalk_sources1, rigetti_xtalk, xtalk_sources3, flammia_wallman_xtalk, google_xtalk}. This paper focuses on crosstalk mitigation in superconducting systems from IBM.
Current IBM systems have gate error rates of 1-2\% per two-qubit operation~\cite{triq}; when affected by crosstalk noise, we observe that gate errors can worsen by an order of magnitude. Through an extensive study, we show that these crosstalk effects can significantly impact the total program reliability. Our goal is to mitigate this error in software through intelligent instruction scheduling. 

Figure~\ref{fig:intro_example} shows the error tradeoffs that influence scheduling decisions. When a pair of simultaneous program operations has high crosstalk, they can be scheduled serially by using control instructions such as barriers. However, naive serialization is harmful. Quantum states are extremely fragile --- the quality of quantum information loses ``coherence'' exponentially with increasing compute time. Because of this exponential decay, current hardware and compilers such IBM Qiskit~\cite{qiskit}, Rigetti Quilc~\cite{quil}, and TriQ~\cite{triq} schedule maximum instructions in parallel. An ideal schedule balances the need for crosstalk mitigation against the need to compute before decoherence. 

Figure \ref{fig:crosstalk_mitigation_flow} shows our crosstalk mitigation approach. Crosstalk mitigation in software requires first, an accurate characterization of crosstalk noise present in the hardware, and second, an intelligent scheduler that uses characterization data to navigate the crosstalk-coherence tradeoff. We advance the state-of-the-art on both fronts.

Our contributions include the following. First, it is known that device characteristics affect compilation quality and program reliability\cite{asplos}. However, measuring all device characteristics (akin to measuring the full process map) is an intractable problem due to exponential scaling.  Therefore, the performance of such devices is typically judged based on a few metrics, such as the gate error rates and the qubit lifetimes, which are collected daily for current QC systems. This paper quantifies the degree to which crosstalk has an important effect on program reliability. 

Second, since measuring crosstalk noise on every pair of simultaneous operations is computationally expensive, (requiring more than 8 hours of machine compute time even for a 20-qubit device), we develop approaches to reduce this overhead. We implement our methods in IBM Qiskit Ignis \cite{qiskit-ignis}, an open source toolbox for device characterization. On \numdevices\ 20-qubit IBM devices, our optimizations reduce characterization time to under 15 minutes.

Third, our evaluation offers insights about crosstalk noise: crosstalk can degrade the error rate of a two-qubit operation up to 11x. The degradation is not static; the effect of crosstalk on a particular gate varies up to 3x over many days. 
On all the three devices in our study, crosstalk noise primarily affects only nearest-neighbor gates. 

Fourth, we develop an instruction scheduler that mitigates the application impact of crosstalk. We model the gate scheduling problem as a Satisfiability Modulo Theory (SMT) optimization and find optimal schedules.
We implement our scheduler in IBM Qiskit Terra \cite{qiskit-terra}, an open-source QC compiler. Using real-system runs on \numdevices\ IBMQ systems, we show that crosstalk mitigation improves the error rate of SWAP circuits by up to 5.6x, geomean 2x over the parallel instruction scheduler previously used by default in IBM systems. Since SWAP operations are the fundamental method of communication in these systems, this large improvement impacts all programs that rely on communication \cite{cross_swap, asplos, triq, mapping2, tannu_qureshi, cgo18, mapping3, wille1}, especially as systems scale up. Our scheduler also improves the loss in cross entropy for QAOA circuits by up to 3.6x compared to the IBM scheduler. In addition, using executions on crosstalk-free regions of the hardware, we empirically verify that the mitigation provided by our approach is near-optimal in practice.


Finally, this work makes the case for software mitigation of crosstalk. This is timely as the trend in quantum computer architecture is moving towards combating crosstalk solely in hardware, either by building more sparsely-connected qubits (such as IBM systems \cite{hexagon_paper_cross}) and/or by disabling simultaneous nearby gates entirely at the hardware level (such as in Rigetti and Google's Bristlecone system~\cite{nasa1, google_xtalk_hw}). Both approaches impose a performance burden when mapping applications to the hardware. Instead, we argue that compilers can better navigate the design tradeoffs. 



%% file: txt/background.tex
\section{QC Background}
\label{sec:background}
\subsection{Principles of Quantum Computing} 
A qubit is the fundamental building block of a QC system. Qubits have two basis states $\ket{0}$ and $\ket{1}$. Unlike classical bits, qubits can also be in superposition, where the state is $\alpha \ket{0} + \beta \ket{1}$, for $\alpha, \beta \in \mathbb{C}, |\alpha|^2+|\beta|^2=1$. When all $n$ qubits in a QC system are in the maximal superposition state ($|\alpha|^2=|\beta|^2=0.5$), the system represents $2^n$ basis states simultaneously, unlike classical systems (non-quantum) which can be in exactly one of the $2^n$ states at any given time.

Instructions or operations in a QC system are termed gates. Gates manipulate information by modifying the complex amplitudes associated with the qubit basis states. The hardware to implement QC gates is designed to apply some dynamic physical interaction to the qubit using a time-dependent set of control signals. For example, in IBMQ systems, gates are implemented by driving the qubits with microwave voltage pulses~\cite{software_gates}. 
Two-qubit Controlled NOT gates are implemented using the cross-resonance effect~\cite{CR1, CR2} where a pulse is applied on control qubit at the resonant frequency of a target qubit. This gate produces entanglement among qubits, which results in non-classical correlated behaviour. 


In a QC application, an algorithm is mapped to gates which execute on a set of appropriately initialized qubits. During execution, qubit states are manipulated and the state space is evolved towards the desired output. At the end of the algorithm, a classical output bitstring can be generated using readout operations which collapse each qubit state's to $\ket{0}$ or $\ket{1}$.

\subsection{Operational Noise in NISQ Systems}
QC systems have spatial and temporal noise variations arising from manufacturing imperfections, imperfections of gate implementation and control, and external interference~\cite{mit_review, triq}. These systems are calibrated frequently to reduce operation noise; during calibrations, error rates are measured using randomized benchmarking~\cite{RB1} and reported for each gate~\cite{ibmq}. 

For the systems used in our study, the error rates for single qubit operations are less than 0.1\%. Error rates for two-qubit CNOT gates range from 0.5-6.5\%, average 1.8\%. Readout error on a single qubit is 4.8\% on average. These error rates indicate the reliability of the operation when it is performed in isolation. QC executions consist of sequences of gates followed by readout, and the errors compound.

While such standalone gate error rates are measured daily, error rates for the simultaneous execution of multiple of these gates have been time-consuming to characterize and therefore are not measured daily. This paper demonstrates that such simultaneous error characterizations are useful, since they can be exploited in the compiler to mitigate the impact of crosstalk. 

%% file: txt/related_work.tex
\section{Related Work and Novelty}
A vast body of prior work exists on quantum circuit optimization to reduce the total number of gates or number of layers in the dependency graph (depth). Refs.~\cite{optimization1, optimization2, optimization3, optimization4} optimize abstract program IR, without considering hardware constraints, while~\cite{mapping1, mapping2, mapping3, mapping4, mapping5, Metodi06schedulingphysical} develop optimizations for mapping programs to hardware qubits to reduce the circuit size or depth. Refs.~\cite{intel1, scheduling_maslov1} use commutation rules to minimize program duration. Ref~\cite{nasa1} considers the case where gates in proximity are disabled from operating simultaneously due to crosstalk, but takes that as disabled in hardware.

Almost all prior work takes it for granted that lower program duration (a.k.a quantum circuit depth) is better, and do not consider crosstalk effects. This is intuitive, since qubits lose their information at an exponential rate as time passes. However, in this work, we show that program duration can be traded off to avoid crosstalk, and thus improve the overall reliability of application executions.

Recently, ~\cite{asplos, tannu_qureshi, triq, rodvanmeter} used hardware characterization data to improve the quality of compilation. They improve the quality of mapping and SWAP insertion using independent gate error rate data measured and published by QC vendors, which does not include crosstalk characterization. Consequently, neither these works nor industrial compilers such as IBM Qiskit~\cite{qiskit}, Rigetti Quilc~\cite{quil} or Google Cirq~\cite{cirq} consider crosstalk effects.

On the hardware side, sparse qubit connectivity \cite{hexagon_paper_cross}, frequency allocation techniques \cite{8614500} and gate implementation approaches \cite{sarah_sheldon_xtalk} have been used to reduce crosstalk. These approaches are complementary to our work and are implemented in the hardware used for our evaluation. 
Hardware scheduling techniques have also been developed. In IBM systems, the default hardware scheduler allows maximum parallelism and aligns all gates to the right to execute them late as possible. Figure \ref{figure:intro_ibm_schedule} shows an example. While this optimization reduces decoherence errors, it does not reduce crosstalk. In Rigetti and Google Bristlecone systems, the hardware scheduler disables simultaneous nearby gates entirely to avoid crosstalk \cite{nasa1, google_xtalk_hw}, irrespective of other hardware or application characteristics. This approach incurs high decoherence error because of excessive serialization. 
Our work proposes the first software technique for crosstalk mitigation and develops an instruction scheduler that serializes instructions to avoid crosstalk, but also balances the need to mitigate decoherence errors.

To the best of our knowledge, this work is the first to evaluate schedule qualities on real quantum systems (with real-world noise characteristics), and the first to improve schedule qualities by considering spatial as well as temporal aspects of the schedule --- that is, which operations should be scheduled when and in proximity to which other operations. Our work is also the first to quantitatively show the extent to which crosstalk effects influence the reliability of programs. 

%% file: txt/designquestions.tex
\section{Crosstalk Mitigation in Software: Design Questions}
\subsection{Background on Crosstalk Sources in Superconducting Systems}
In superconducting systems, crosstalk can occur for several reasons. One type of crosstalk is due to the hardware necessary to couple pairs of qubits for two-qubit operations. There is a tradeoff between the strength of these couplings and the known, but unwanted, crosstalk they generate. In IBM devices, each qubit is connected to a few other nearest-neighbor qubits through fixed-frequency microwave resonators resulting in an always-on coupling. In Figure \ref{fig:machine_figs}, each CNOT gate (edges) corresponds to one resonator. Because of the always-on nature of the coupling, when a control pulse is driven on one of the qubits, the resonator can propagate an unwanted drive to neighboring qubits and corrupt their state. This effect is particularly acute for nearby qubits that have similar resonant frequencies. If multiple nearest neighbor or next nearest neighbor qubits have overlapping resonant frequencies, driving a qubit can lead to unwanted state changes on other qubits. Despite meticulous efforts to mitigate crosstalk in QC hardware \cite{sarah_sheldon_xtalk, hexagon_paper_cross}, crosstalk noise is present in real devices \cite{robin_xtalk, flammia_wallman_xtalk}.

\subsection{Characterizing Crosstalk Noise Through Randomized Benchmarking}
To mitigate crosstalk noise in software, we must first characterize the hardware. For example, in Figure \ref{fig:intro_example}, to quantify the impact of crosstalk for the gates $g_1$ and $g_2$ executing in parallel, we have to measure the crosstalk noise for the corresponding hardware gates CNOT 0,1 and CNOT 2,3. To accomplish this, the error rate of CNOT 0,1 and CNOT 2,3 can be measured independently, without invoking any other gate. Then, the error rate of CNOT 0,1 and CNOT 2,3 can be measured simultaneously, by invoking them in parallel. If the simultaneous error rates are much higher than the independent error rates, crosstalk exists between the two gates.

Such measurements can be performed using Randomized Benchmarking (RB), a standard procedure for measuring gate error rates, which is used in IBM systems \cite{RB1, RB2, RB3}. To measure the error rate, a single invocation of a gate is not enough. For CNOT error measurement, RB uses multiple random circuits, each having multiple invocations of the CNOT composed along with random single qubit operations. By executing these circuits on the hardware and fitting the results to a theoretical model, the error rate is estimated. For a gate $g_i$, we denote the error rate measured without invoking any other gate in the system as the {\em independent error rate} $E(g_i)$ and the error rate of $g_i$ measured simultaneously with $g_j$ as a {\em conditional error rate} $E(g_i|g_j)$. Simultaneous RB (SRB) on a pair of gates $g_i$ and $g_j$ yields both $E(g_i|g_j)$ and $E(g_j|g_i)$. When a gate $g_i$ has crosstalk interference with $g_j$, we expect $E(g_i|g_j)$ to be higher than $E(g_i)$. 

While independent error rates are available from daily calibration data, conditional errors are not. To measure conditional error rates for a device, we have to perform SRB experiments between every pair of CNOT gates that can be driven in parallel i.e., CNOT pairs such as CNOT 0,1 and CNOT 2,3 that do not share a qubit. For \ibmqpoughkeepsie\, this approach requires 221 pairs of SRB experiments. Each such SRB experiment requires multiple runs with different random gate lengths (to get the final curve fit to the theoretical model) and each data point on the curve requires multiple trials because of noisy operations. With 100 random sequences per SRB, and 1024 trials per sequence, this baseline method requires 22.6M executions and over 8 hours of computation at current execution rates. Since QC systems have highly variable noise properties \cite{triq}, daily crosstalk measurement (similar to daily gate error measurement which is already performed by IBM) will consume over a third of a device's total lifetime.

Therefore, we ask: 
{\em How can we perform crosstalk characterization experiments efficiently across the full device? Can we exploit the physical properties of the device to reduce the number of experiments? What crosstalk measurements are useful for mitigation in software?} 




\subsection{Mitigating Crosstalk by Instruction Scheduling}
To avoid crosstalk, a compiler can choose to serialize the interfering operations. However, serialization can lead to decoherence errors. On IBM systems, coherence times on individual qubits range from 10-100 microseconds \cite{triq} --- when a program executes for 50 microseconds on the best qubit with 100us coherence, it is 60\% likely that the state is corrupted. To mitigate this dramatic loss in reliability from decoherence, the compiler should parallelize instructions as much as possible.

We ask: 
 {\em How can a compiler optimize the two conflicting objectives of serializing instructions to avoid high crosstalk and parallelizing instructions to avoid decoherence? How much does crosstalk-adaptivity matter for parallel operations on superconducting devices?}

%% file: txt/body.tex
\section{Reducing the Crosstalk Characterization Overhead}\label{sec:xtalk_opt}
\begin{figure*} [t]
    \centering
    \subfloat[\ibmqpoughkeepsie]{
    \includegraphics[scale=0.45]{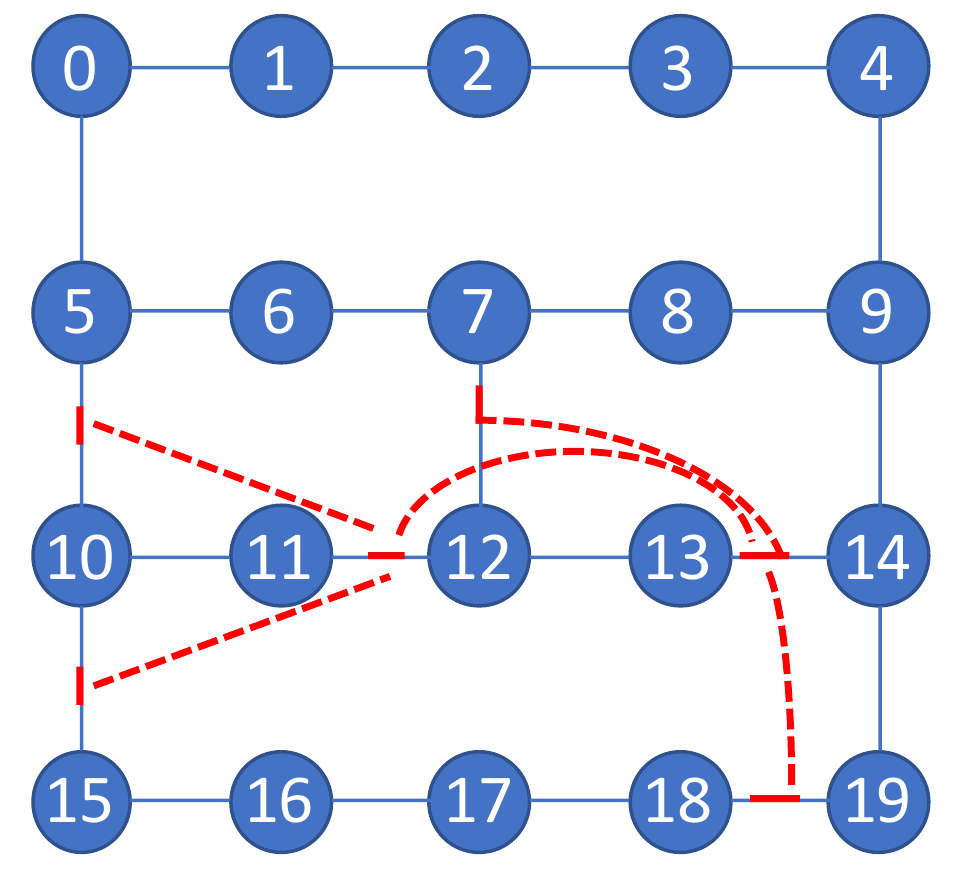}
    \label{fig:poughkeepsie_xtalk}
    }\quad
    \subfloat[\ibmqsystemone]{
    \includegraphics[scale=0.45]{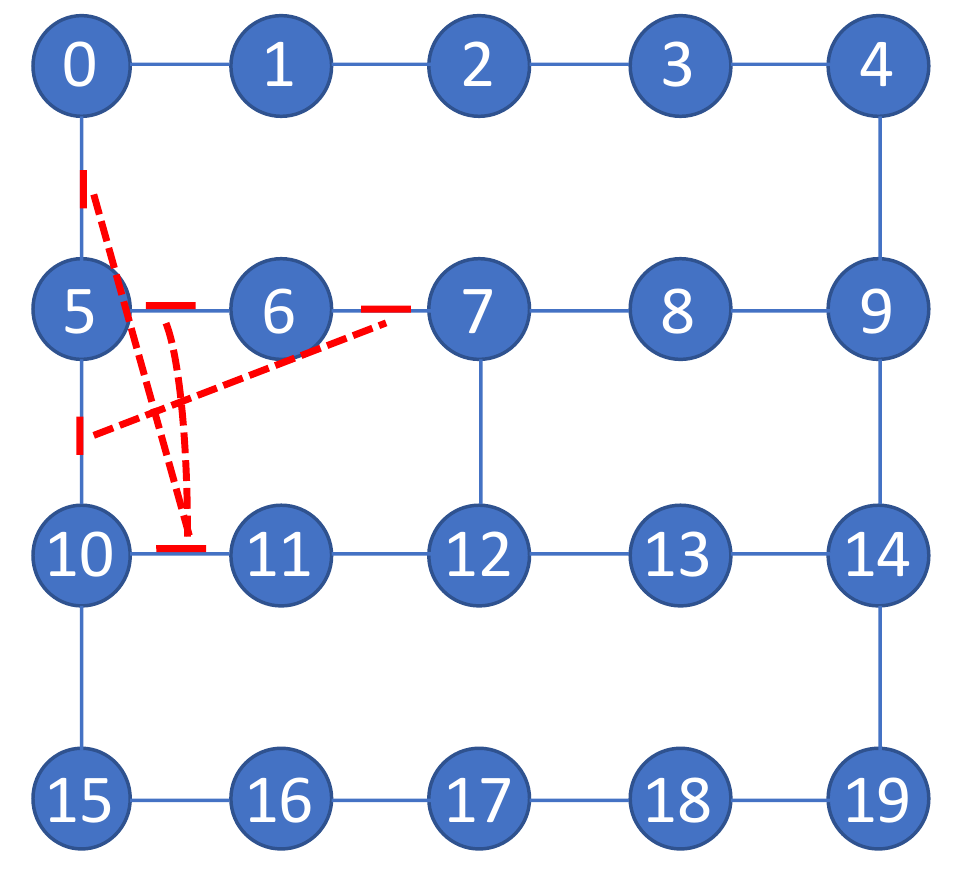}
    }\quad    
    \subfloat[\ibmqboeblingen]{
    \includegraphics[scale=0.45]{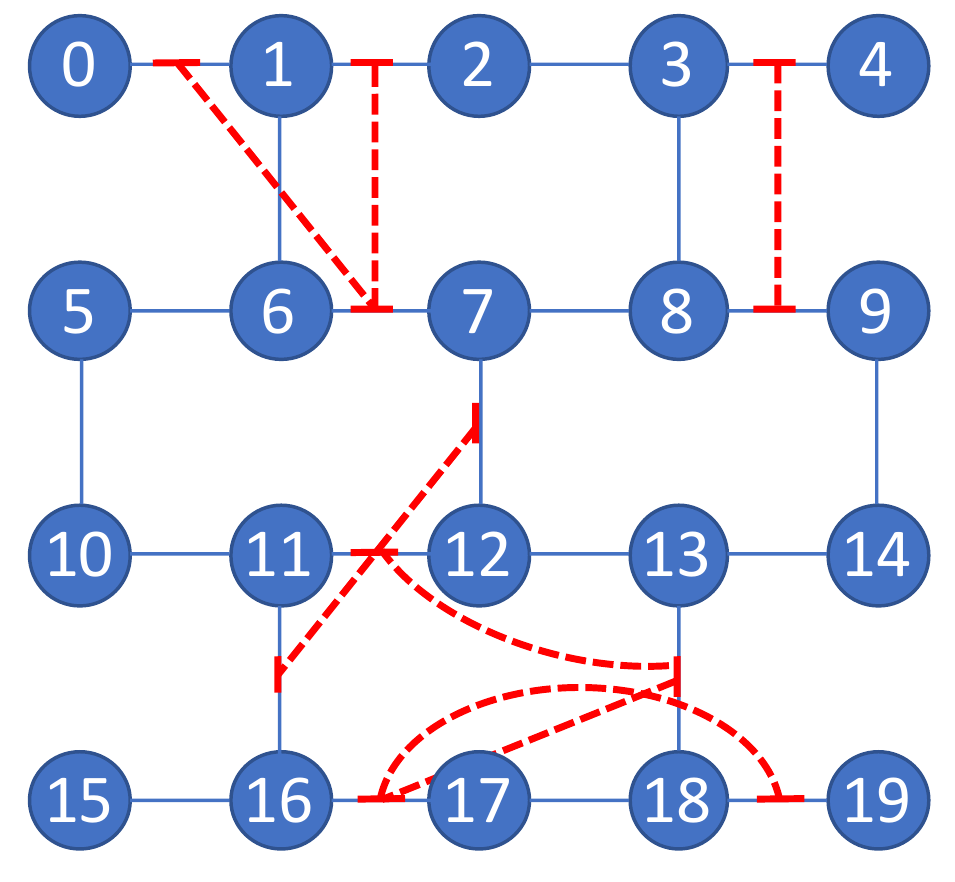}
    }
    \caption{Crosstalk measurement results for the three systems. The nodes are indexed by qubit id. The edges indicate two-qubit CNOT gates. Note that the number of connections is less than a regular 2D grid. For all CNOT gates, the independent gate error rate is at most $5\%$. We performed simultaneous RB experiments on all pairs of CNOT operations in the hardware, one pair at a time. SRB experiment on a pair $g_i$ and $g_j$ gives $E(g_i|g_j)$ and $E(g_j|g_i)$. We illustrate the data by drawing {\color{red} red} dashed edges to indicate {\em high crosstalk} gate pairs i.e., all gate pairs $g_i$, $g_j$ for which the conditional error rate is much higher than the independent error rate. For this plot we selected CNOT pairs such that $E(g_i|g_j) > 3*E(g_i)$ or $E(g_j|g_i) > 3*E(g_j)$. For example, on \ibmqpoughkeepsie, CNOT 10, 15 has an independent error rate of 1\% and conditional error rate of 11\% with CNOT 11, 12.}
    \label{fig:machine_figs}
\end{figure*}
\hide{
\begin{figure*}[t]
    \centering
    \subfloat[\ibmqpoughkeepsie]
    {
    \includegraphics[scale=0.35]{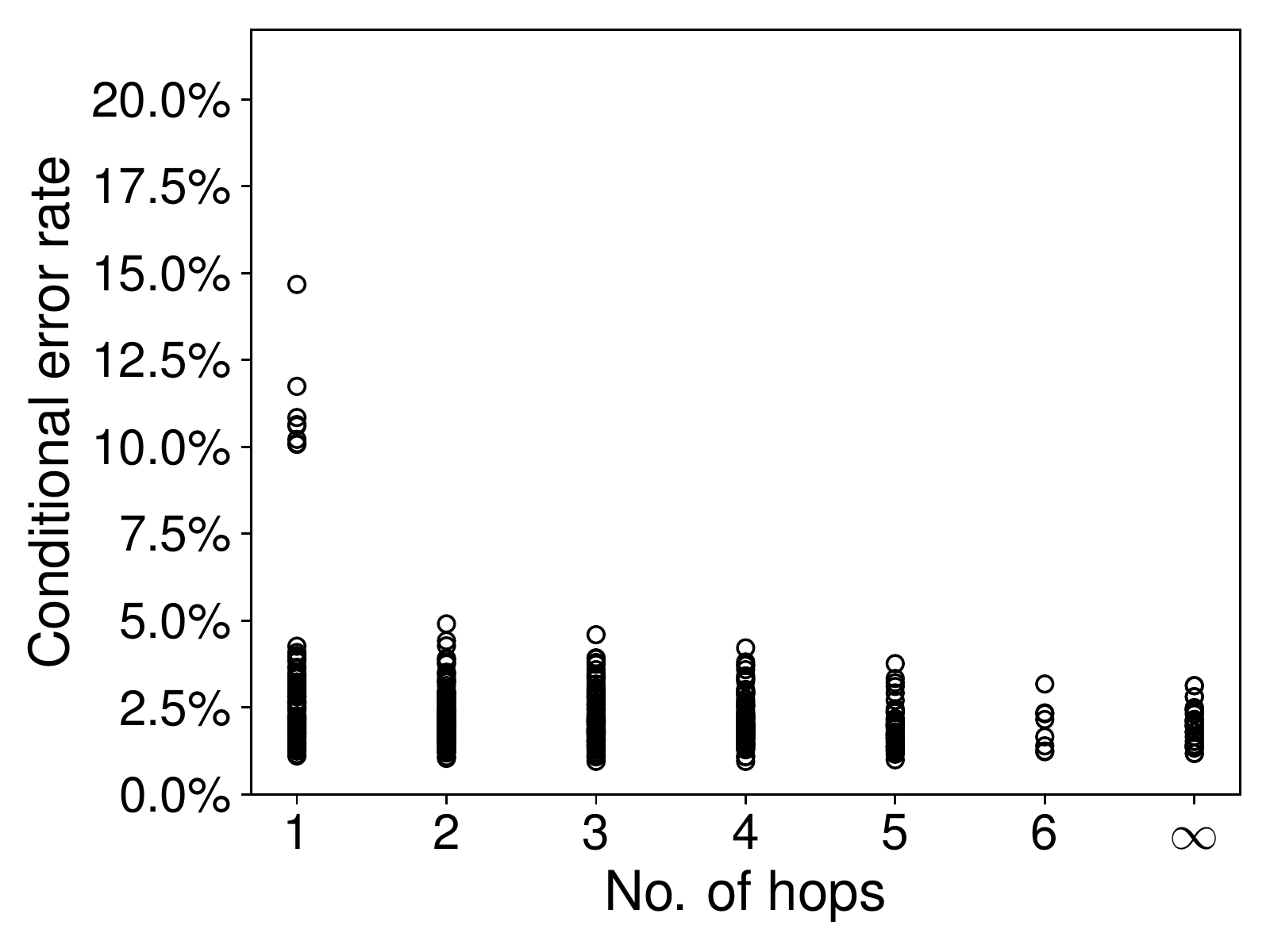}
    \label{fig:pougkeepsie_box_plot}
    }
    \subfloat[\ibmqsystemone]
    {
    \includegraphics[scale=0.35]{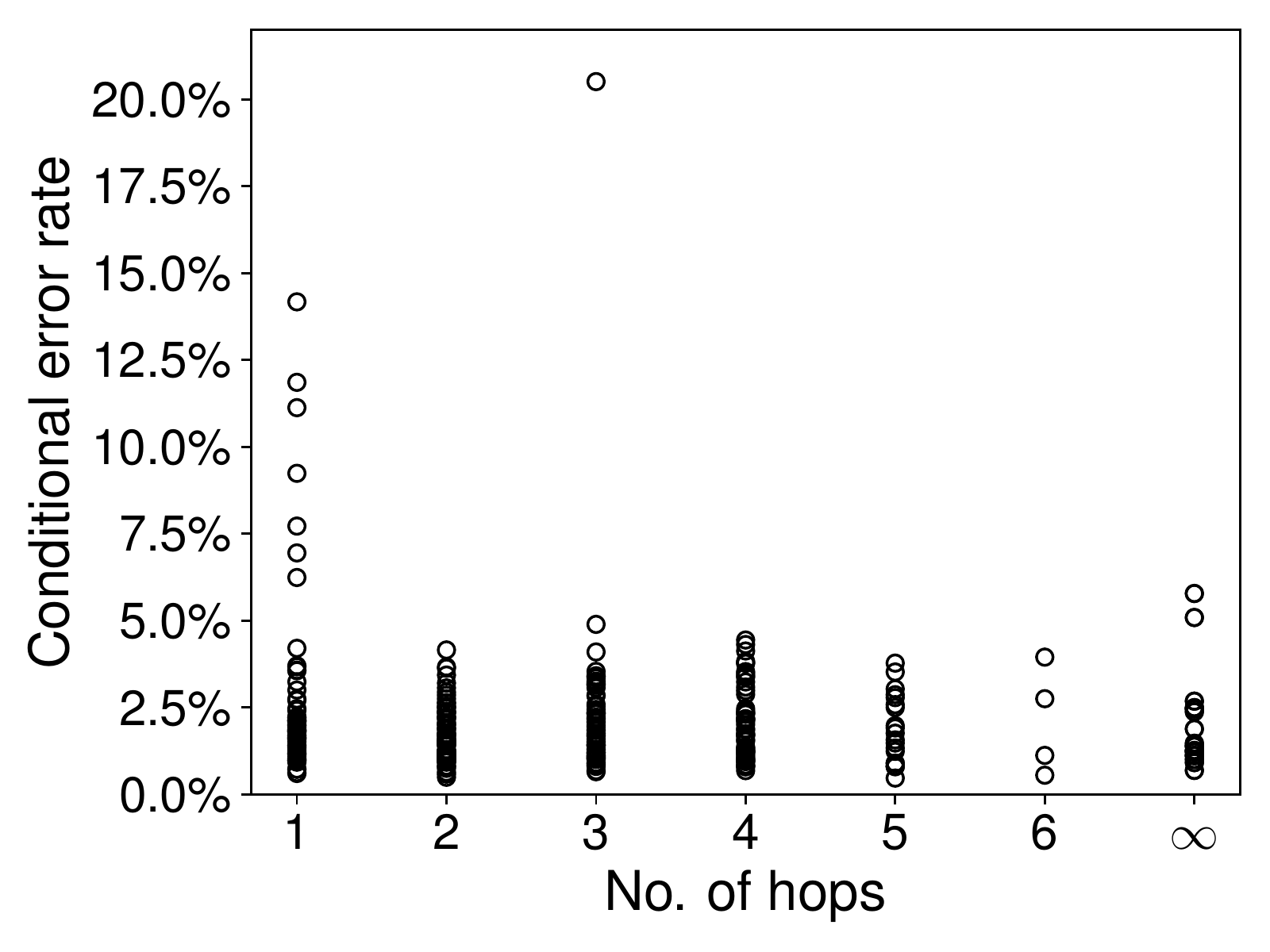}    
    \label{fig:systemone_box_plot}
    }
    \subfloat[\ibmqboeblingen]
    {
    \includegraphics[scale=0.35]{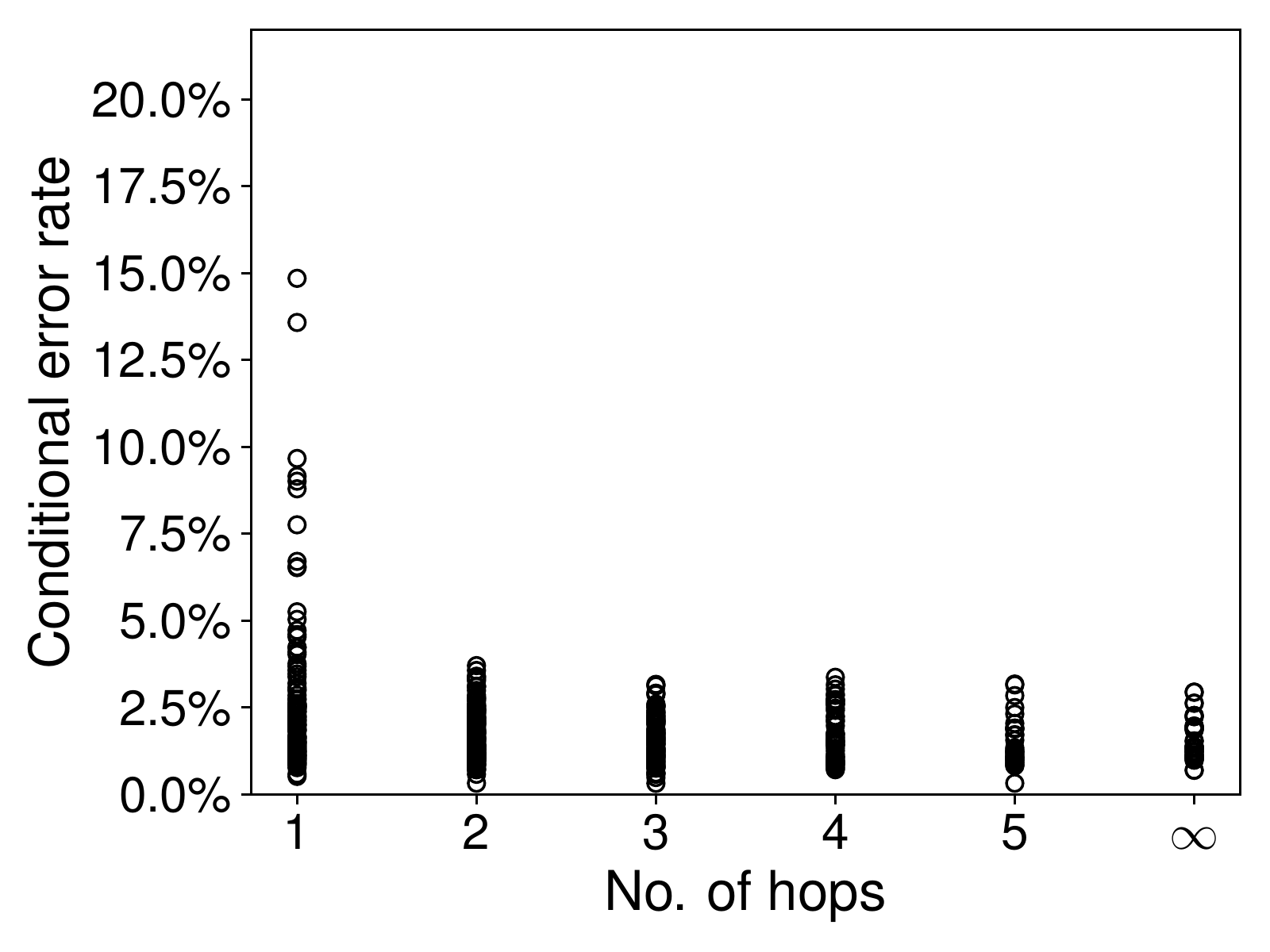}    
    \label{fig:boeblingen_box_plot}
    }
    \caption{Conditional error rates between gate pairs at different distances. Lower error rate is better. We performed simultaneous RB experiments on all pairs of CNOT operations in the hardware, one pair at a time. SRB experiment on a pair $g_i$ and $g_j$ gives $E(g_i|g_j)$ and $E(g_j|g_i)$. At hop count $k$, the figure shows the scatter plot of these conditional error rates for gate pairs separated by $k$ hops. For easy comparison, at distance $\infty$ we plot the independent, crosstalk-free error rates for each gate. In Figure (a), at hop count 1 there are several gate pairs which have conditional error rates higher than 7.5\%., which is much higher than the maximum Y-axis value of the scatter plot at distance $\infty$, indicating crosstalk effects at 1 hop distance.
    On these 3 devices, crosstalk noise from a gate primarily influences only gates which are at 1 hop distance. }
    \label{fig:box_plots}
\end{figure*}
}
\begin{figure}[t]
    \centering
    \includegraphics[scale=0.5]{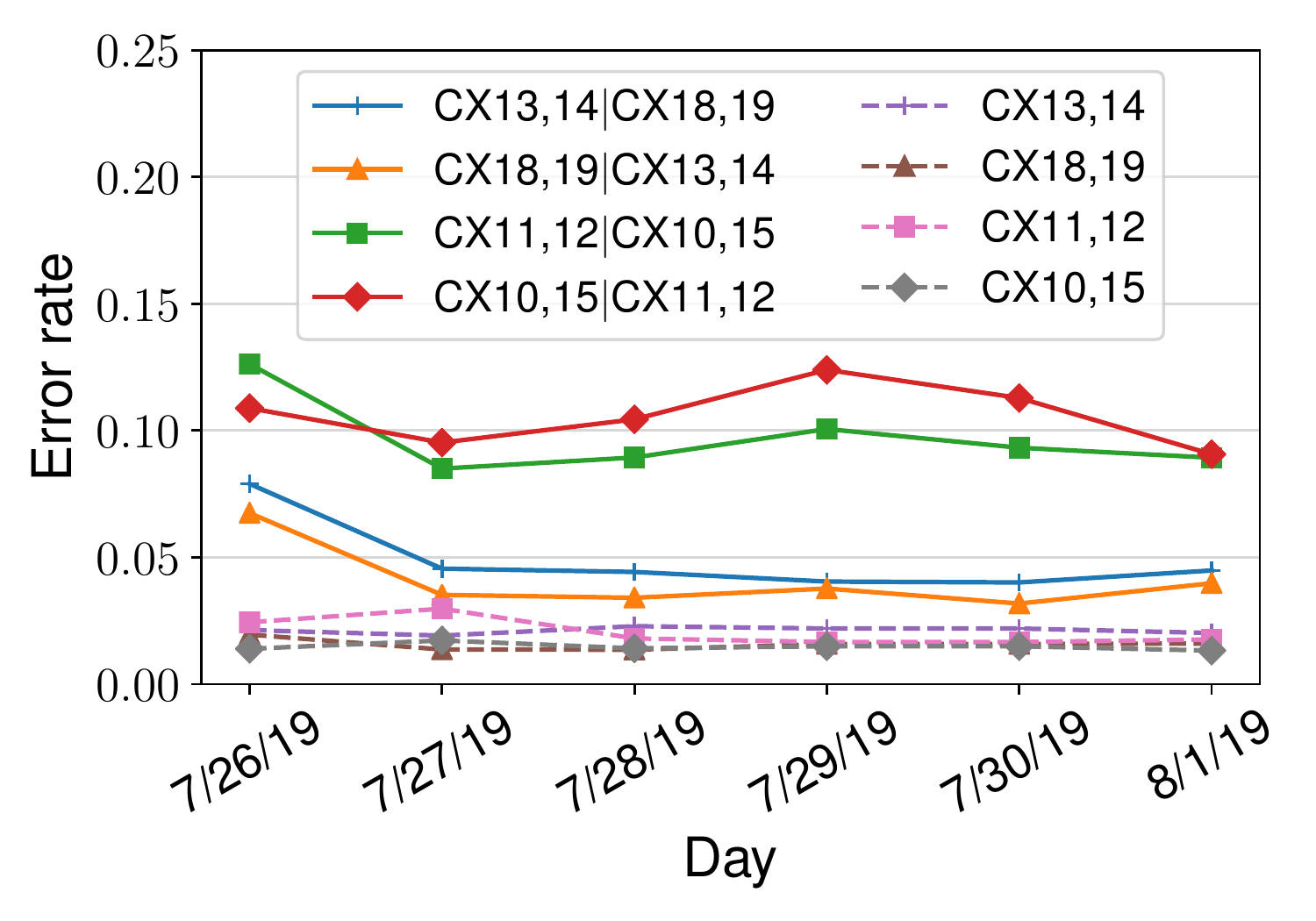}
    \caption{Daily variations of crosstalk noise in \ibmqpoughkeepsie. Lower error rate is better. Conditional error rates of a gate, say $E(CX13,14|CX18, 19)$ are much higher than the independent error rate throughout the experiment week. The conditional error rates vary up to 2x on this machine, and up to 3x across devices.}
    \label{fig:pougkeepsie_daily}
\end{figure}
We first present detailed characterization results on \numdevices\ IBM systems. Using insights from our characterization, we propose optimizations to reduce the characterization overhead.

\subsection{Characterization Results on IBMQ Systems}
Figure \ref{fig:machine_figs} illustrates the crosstalk measurements for the \numdevices\ systems. Across the gate pairs, crosstalk noise increases the gate error rates up to 11x. For \ibmqpoughkeepsie, high crosstalk errors occur only in 5 gate pairs, a small fraction of the overall number of gate pairs (221). In each interfering pair, the two gates are separated by 1 hop i.e, the shortest path from one gate to the other is of length 1, which is the expected behavior from device design.


Figure \ref{fig:pougkeepsie_daily} shows daily variations of the error rates for \ibmqpoughkeepsie. Conditional error rates for a gate vary up to 2x for \ibmqpoughkeepsie, and up to 3x on the other two systems (not shown). Even though the absolute error rates vary, the set of high crosstalk gate pairs tends to remain the same across days.

\subsection{Our Optimizations}
To mitigate crosstalk through instruction scheduling, we require accurate characterization data. Since crosstalk noise has spatio-temporal variations, it should be characterized daily to supply correct inputs to the compiler, similar to how gate errors and coherence times are measured daily on IBM systems. Towards this, we wish to reduce the number of experiments required to measure conditional gate error rates.

In the previous section, we measured conditional error rates for {\em every pair} of CNOT gates that can be driven in parallel. For \ibmqpoughkeepsie\ this approach requires 221 pairs of SRB experiments and over 8 hours of real-system compute time on the QC device. \emph{All these experiments are performed on the hardware, not in simulation}. At face value, and without knowledge of the spatio-temporal behavior of crosstalk, this means that to enable compiler-level mitigation of crosstalk we must run this expensive characterization step daily. However, through a series of optimizations, we can reduce the characterization overhead. 

\paragraph{Optimization 1: Characterize only 1-hop pairs.} It is sufficient to perform SRB experiments on gate pairs which are separated by 1 hop since on our devices, crosstalk noise from a gate is significant only at 1 hop distance (see Figure \ref{fig:machine_figs}). This is the expected behavior from device design, since qubits are dispersively coupled, i.e., the ratio of coupling strength to detuning is much less than one. For each additional hop the effective coupling is suppressed by this dispersive factor. However, device packaging imperfections have been seen to introduce longer range crosstalk effects in some older systems \cite{flammia_wallman_xtalk,google_xtalk,rigetti_xtalk}.

\paragraph{Optimization 2: Parallelize SRB experiments of multiple gate pairs.} Next, given the above observation about lack of long-range crosstalk, we can efficiently parallelize crosstalk measurements across several gate pairs. When two pairs are separated by two or more hops, their SRB measurements can be performed in parallel. For example, in \ibmqpoughkeepsie, we can perform crosstalk measurement for the pairs (CNOT 0,1 | CNOT 2,3), (CNOT 6,7 | CNOT 8,9) and (CNOT 16,17 | CNOT 18,19) in the same experiment since each pair is at least 2 hops away from any other pair. 

To efficiently parallelize SRB experiments, we can model the problem as an instance of bin packing. Given a set of $n$ gate pairs on which SRB measurements are required, we use a randomized first fit heuristic to pack the pairs into a small number of experiments. The heuristic iteratively builds a set of bins, with each bin corresponding to an experiment. Initially, there is only one empty bin. The heuristic iterates through the gate pairs and places each pair in the first {\em compatible} bin. A pair $(g_i, g_j)$ is compatible with a bin if all pairs $(g_k, g_l)$ in the bin are at least $k$ hops away. For example in \ibmqpoughkeepsie, with k=2, the pair (CNOT 16,17 | CNOT 18,19) is compatible with a bin which contains the pair (CNOT 6,7 | CNOT 8,9); it is not compatible with a bin which contains the pair (CNOT 11,12 | CNOT 13,14). When no existing bin is compatible, a new bin is created. All gate pairs are partitioned into a set of bins in this manner. We repeat the algorithm multiple times by shuffling the list of gate pairs randomly and select the partitioning with the minimum number of bins. We perform SRB experiments in parallel for all gate pairs that belong to the same bin.

\paragraph{Optimization 3: Characterize high crosstalk pairs only.} Finally, from our characterization data over several days for these devices, the set of high-crosstalk pairs remains relatively stable across days (see Figure~\ref{fig:pougkeepsie_daily}). This is due to the structural nature of crosstalk pairs, and compared to gate errors, it is less prone to drift or regular changes. Hence, we can optimistically restrict our daily measurements on these pairs, and periodically, say once every few days, characterize the remaining 1 hop pairs. 

Combining all optimizations, we can reduce the characterization time for the \numdevices\ systems to under fifteen minutes. After  characterization, the data can be used by all compilation jobs in this period to improve their output.



%% file: txt/scheduler.tex
\section{Crosstalk Mitigation Through Instruction Scheduling: Overview}
The input to our scheduler is a hardware-compliant program IR i.e., the program qubits are mapped to the hardware qubits and the IR includes the necessary SWAP instructions required to respect connectivity constraints. Figure \ref{fig:intro_example_prog} illustrates such a program IR for the example machine in Figure \ref{fig:intro_example_machine}. In our implementation using IBM Qiskit, we obtain such IR by invoking existing passes for mapping and SWAP insertion. The scheduler uses crosstalk characterization data along with machine calibration data (independent error rates, coherence time, gate duration) to determine a start time for each gate.

We pose the gate scheduling program as a constrained optimization problem to be solved by a Satisfiability Modulo Theory (SMT) solver \cite{z3, omt_z3}. The optimization has variables and constraints which express program information and hardware error information. The variables in the optimization include the start time and error rate for each gate. We use gate dependency constraints to specify that the schedule should preserve program data dependencies. 

To model the effect of crosstalk, we should determine the error rate of a gate based on the program schedule. When a gate does not overlap in time with other operations, its error rate is set using crosstalk-free independent error rates. When the gate overlaps with other operations, the error rate is based on conditional error rates with the overlapping operations. For each gate, we determine the set of overlapping operations based on the IR dependencies. The subsets of this set are the various gate overlap scenarios and are used to set the appropriate conditional error rates for a gate.

To model the effect of decoherence, we associate a {\em lifetime} variable with every qubit. The lifetime is the time elapsed between the first operation and the last operation on the qubit. We associate a decoherence error rate variable with a qubit which is computed as an exponential penalty on the lifetime, normalized by the coherence time of the qubit. Thus, when the lifetime increases, the decoherence error rate increases. 

The objective function captures the tradeoff between instruction serialization for crosstalk mitigation and parallelization for decoherence mitigation. We minimize the product of gate error rates (which are influenced by crosstalk) and the qubit error rates (which are based on decoherence). 
When the optimizer serializes two gates which have high crosstalk, the gate error rate terms reduce and the decoherence terms increase. Similarly, when the gates are executed in parallel, the gate error terms increase and the decoherence terms reduce. Minimizing the objective over the entire program allows us to find the optimal schedule which mitigates crosstalk while also balancing the errors from decoherence.

Finally, to implement the schedule and enforce gate orderings, we use a post-processing step to insert control instructions in the form of barriers.
We call our scheduler \xtalksched. We compare its performance to two baselines \seriessched\ and \parsched. These variants are discussed in Table \ref{tab:sched_variants}.

\begin{table}
\centering
\footnotesize
\begin{tabular}{|l|l|l|}
\hline
Algorithm   & Objective                                                                     & Method                                                                                                                                                \\ \hline \hline
\seriessched & Mitigate crosstalk                                                            & \begin{tabular}[c]{@{}l@{}}Schedule all instructions\\ serially\end{tabular}                                                                          \\ \hline
\parsched    & Mitigate decoherence                                                          & \begin{tabular}[c]{@{}l@{}}Schedule maximum \\ instructions in parallel. \\ Current state-of-the-art,\\ used in Qiskit \cite{qiskit}, \\ Quilc \cite{quil} and TriQ \cite{triq}\end{tabular}                          \\ \hline
\xtalksched  & \begin{tabular}[c]{@{}l@{}}Mitigate crosstalk \\ and decoherence\end{tabular} & \begin{tabular}[c]{@{}l@{}}SMT optimization with \\ crosstalk characterization \\ data. $\omega$: crosstalk weight\\  factor. (Section 6, 7)\end{tabular} \\ \hline
\end{tabular}
\caption{List of schedulers used in our evaluation.}
\label{tab:sched_variants}
\end{table}

\section{Instruction Scheduling: Optimization Details}
\subsection{Variables}
Let $Q$ be the set of qubits and $G$ be the set of gates in the IR. For each gate $g \in G$, the start time is denoted by ($g.\tau$), duration by ($g.\delta$), and error rate by ($g.\epsilon$). To denote data dependencies between two operations, we use a binary relation $>$ on the gates. For two operations $g_j > g_i$ if $g_j$ depends on $g_i$. In addition to these variables, for each qubit $q$ in the program, we create a coherence error rate variable $q.\epsilon$.


\subsection{Constraints}
\hfill\\
{\bf Data dependency constraints: }
If two gates $g_i$ and $g_j$ operate on the same qubit, and $g_j$ uses the output of $g_i$, $g_j$ should start only after $g_i$ finishes.
Such dependencies can be enforced by the following constraint.
\begin{gather}
    \forall g_i, g_j \in G : g_j > g_i \Rightarrow g_j.\tau \ge g_i.\tau + g_i.\delta 
\end{gather}
For example, for Figure \ref{fig:intro_example_prog}, the constraint $g_1 > g_0.\tau + g_0.\delta$ expresses the data dependency between $g_0$ and $g_1$.

Gate duration information is available to the compiler either from machine documentation or from calibration data and is used to set the duration variables $\delta$.

{\noindent \bf Gate error constraints:}
These constraints set crosstalk dependent error rates for each two-qubit gate. We don't consider conditional error rates based on single qubit gates because their error rates are 10x better than CNOT error rates on current systems~\cite{triq}.
For each gate $g_i$ denote by $CanOlp(g_i)$, the set of all operations that can overlap with it. This set can be computed by finding each $g_j$ that is neither an ancestor nor a descendent of $g_i$ in the program dependency graph specified by the IR. In Figure \ref{fig:intro_example_prog}, $CanOlp(g_2) = \{g_1, g_3\}$. $g_0$ is not considered because it is a single-qubit gate. We prune this set further to only include gates which have high conditional error rates, which in our systems are at 1 hop distance from $g_i$.

For each gate $g_j \in CanOlp(g_i)$, we create an overlap indicator $o_{ij}$, which tracks whether $g_i$ and $g_j$ overlap in the schedule. $o_{ij}$ is set using the following constraint.
\begin{gather}
    o_{ij} = (g_j.\tau \le g_i.\tau + g_i.\delta \land g_i.\tau \le g_j.\tau + g_j.\delta) 
\end{gather}

How can we set the gate error rates using the overlap indicators? Consider $g_2$ in Figure \ref{fig:intro_example_prog}. Since $g_2$ can overlap with $g_1$ and $g_3$, there are 4 possible scenarios: both $g_1$ and $g_3$ don't overlap with $g_2$, only $g_1$ overlaps with $g_2$, only $g_3$ overlaps with $g_2$, and both $g_1$ and $g_3$ overlap with $g_2$. For each case, we set error rates using the following constraints. 
\begin{align}
\neg o_{12} \land \neg o_{13} &\Rightarrow g_2.\epsilon = E(g_2) \label{egc1} \\
\ o_{12} \land \neg o_{13} &\Rightarrow g_2.\epsilon = E(g_2|g_1)  \label{egc2}\\
\neg o_{12} \land  o_{13} &\Rightarrow g_2.\epsilon = E(g_2|g_3) \label{egc3}\\
o_{12} \land  o_{13}& \Rightarrow g_2.\epsilon = max\{E(g_2|g_1), E(g_2|g_3)\}\label{egc4}
\end{align}

In constraint \ref{egc1}, the error rate is the independent error rate of $g_2$, since it doesn't overlap with the other two gates. In constraint \ref{egc2}, the error rate is the conditional rate of $g_2$ with $g_1$. In constraint \ref{egc4}, when both gates overlap with $g_2$, crosstalk may arise from both gates. But, in order to conservatively serialize gates, we only consider crosstalk from the worst gate, and take the maximum error rate over the two overlapping gates. (We have not observed significant worsening of errors from simultaneous execution of triplets of gates).


We generalize these constraints as follows. To set the error rate for $g_i$, we enumerate all possible overlap scenarios by considering the powerset\footnote{The powerset of a set S is the set of all subsets of S, including the empty set and S itself. The cardinality of the powerset is $2^{|S|}$.} of $CanOlp(g_i)$. For each nonempty subset $Olp_k$ in the powerset, we denote the complement by $NotOlp_k$ i.e., $NotOlp_k = CanOlp(g_i)\setminus Olp_k$ and we add the following constraint.
\begin{gather}
    \bigwedge\limits_{g_j \in Olp_k} o_{ij} \bigwedge\limits_{g_j \in NotOlp_k} \neg o_{ij} \Rightarrow g_i.\epsilon = \max_{g_j \in Olp_k} E(g_i|g_j)
\end{gather}
In other words, when the gates in the set $Olp_k$ overlap with $g_i$, and the gates in the set $NotOlp_k$ don't overlap with $g_i$, the constraint sets the error rate to be the maximum conditional error rate over the overlapping gates.

For the empty subset in the powerset, we add the following constraint to account for the case when none of the gates overlap with $g_i$. 
\begin{gather}
    \bigwedge\limits_{g_j \in CanOlp(g_i)} \neg o_{ij}  \Rightarrow g_i.\epsilon =  E(g_i)
\end{gather}
Although there are $2^{|CanOlp(g_i)|}$ constraints for each gate, in practice the size of the set will not be large because it includes only overlapping gates with high conditional error rates. As Figure \ref{fig:machine_figs} shows, this is small for our systems.

{\noindent \bf Decoherence error constraints:}
These constraints track the decoherence errors on each qubit in the program. They use coherence time measurements available from daily machine calibration.

Exponential state decay in qubits can occur in two ways: $T_1$ time for the state $\ket{1}$ to decay to $\ket{0}$ and $T_2$ time for a superposition state $(\ket{0}+\ket{1})/\sqrt{2}$ to decay to either $\ket{0}$ or $\ket{1}$. These are relaxation and dephasing respectively. We use the term decoherence to refer to both these effects. $T_1$ and $T_2$ values are reported for each hardware qubit during daily calibration.  If a program performs computation for time $t$ on a qubit, the probability of error from $T_1$ losses is proportional to $1-e^{-t/T_1}$, and the probability of error from $T_2$ losses is proportional to $1-e^{-t/T_2}$. When $t$ increases, the error rate increases exponentially.

We set the decoherence error rate for a qubit $q_i \in Q$ by computing the {\em lifetime} of the qubit in the schedule. The lifetime $q_i.t$ is the difference between the finish time of $q_i$'s last gate $L(q_i)$ and the start time of $q_i$'s first gate $F(q_i)$. Current QC systems are typically limited by $T_1$ errors, but on some qubits, $T_2$ times can be much lower than $T_1$ because of noise fluctuations. To consider such cases, we set the maximum available compute time $q_i.T$ as the minimum of $T_1$ and $T_2$ values of the qubit. We set the decoherence error on a qubit as follows.
\begin{gather}
    q_i.t = L(q_i).\tau + L(q_i).\delta - F(q_i).\tau \label{lifetime_constraint}\\
    q.\epsilon = 1-e^{q_i.t/q_i.T}\label{coherence_constraint}
\end{gather}
Although this constraint performs exponentiation over an optimization variable, in the next section we show that it can be expressed as a linear term. 

{\noindent \bf IBMQ-specific constraints: }
Using Qiskit at the circuit level, we can enforce control dependencies only using barrier instructions. Therefore, any schedule where two gates partially overlap cannot be enforced using the circuit-level ISA~\cite{openqasm1}\footnote{Recent versions of Qiskit and IBMQ systems provide a pulse-level abstraction for more fine-grained control of systems~\cite{openpulse_arxiv}}. For each gate $g_i$, and for $g_j \in CanOlp(g_i)$ we enforce that the two gates can either be scheduled without any overlap or such that one of them happens fully within the duration of the other.
\begin{gather}
    (g_i.\tau + g_i.\delta < g_j.\tau) \lor (g_j.\tau + g_j.\delta < g_j.\tau) \lor \\ 
    ((g_i.\tau + g_i.\delta < g_j.\tau + g_j.\delta) \land (g_i.\tau > g_j.\tau)) \lor \\
    ((g_j.\tau + g_j.\delta < g_i.\tau + g_i.\delta) \land (g_j.\tau > g_i.\tau))
\end{gather}

In current IBMQ systems the hardware control forces all readout operations to occur simulateneously at the end. Therefore, all gates are right-justified and scheduled from the end. This affects the qubit lifetime variables in our optimization. We model this behavior with a constraint that equates the start times of all readout operations in the program.

\subsection{Objective Function}
Ideally, to minimize both gate errors from crosstalk and decoherence errors we can set the objective as,
\begin{gather}
     \min\Big(\underbrace{\prod_{\forall g \in G} (g.\epsilon)}_{\text{Gate errors (crosstalk)}}
     \underbrace{\prod_{\forall q \in Q} (q.\epsilon)}_{\text{Decoherence error}}\Big).
\end{gather}
The first term minimizes the product of the gate errors and the second term minimizes the product of decoherence errors. Since the SMT solver requires linear operations, we can minimize the logarithm of the objective to get a linear function.
\begin{gather}
     \min\Big(\sum_{\forall g \in G} (\log g.\epsilon) +  \sum_{\forall q \in Q} (\log q.\epsilon)\Big)
\end{gather}
By substituting the definition for $q.\epsilon$ from constraint \ref{coherence_constraint}, we can re-write the objective as follows.
\begin{gather}
     \min\Big(\sum_{\forall g \in G} (\log g.\epsilon) -   \sum_{\forall q \in Q} (q.t/q.T)\Big)
\end{gather}
In this form the objective function clearly shows the crosstalk-coherence tradeoff. When gates are serialized to reduce crosstalk errors, the first term reduces and the second term increases, and vice versa when gates are parallelized. 

Finally, to test the relative importance of crosstalk and decoherence errors, we consider a weighted objective where a crosstalk weight factor $\omega \in [0,1]$ is applied to the gate error rate terms. 

\begin{gather}
     \min\Big(\omega\sum_{\forall g \in G} (\log g.\epsilon) -   (1-\omega)\sum_{\forall q \in Q} (q.t/q.T)\Big)
\end{gather}

To compute the optimal schedule for a program, we first use Qiskit's passes to generate the program IR and map it to the hardware. The mapped program IR is used to create the optimization problem using this objective along with data dependency, gate error and decoherence error constraints. These constraints make the $g.\epsilon$ and qubit lifetime $q.t$ variables dependent on the gate schedule. The gate schedule produced by the optimization is post-processed to generate executable code with the barriers necessary to enforce the optimal gate orderings. We call this algorithm \xtalksched. 



%% file: txt/expt.tex
\section{Experimental Setup}
\subsection{Crosstalk Characterization Implementation}
We implemented the crosstalk characterization methods using IBM Qiskit Ignis version 0.2.0 \cite{qiskit-ignis}, an open-source framework for error characterization. For CNOT error characterization, RB applies random sequences of $m$ two-qubit Clifford gates which are constructed from single qubit gates and multiple invocations of the CNOT. The final gate is the inverse of the previous gates so that the sequence should return to the original state. By measuring the final state as a function of $m$ and fitting to a theoretical model, one can extract the error rate per Clifford. To extract the CNOT error rate, the Clifford error rate is divided by the by the number of CNOTs per Clifford (optimally 1.5). This assumes the single qubit gate error is negligible and gives an upper bound on CNOT error. To measure a CNOT gate's error rate independently, we apply standard two-qubit RB \cite{qiskit-ignis}. To measure the error rates for two CNOTs simultaneously, we apply simultaneous two-qubit RB (SRB) \cite{SRB}. In each SRB experiment, we used 100 random sequences, with up to 40 Clifford gates per sequence and performed 1024 trials per sequence.

\subsection{Instruction Scheduler Implementation and Baselines}
 We implement our instruction scheduler \xtalksched\ as a compilation passes in IBM Qiskit Terra version 0.8.2 \cite{qiskit-terra}, an open-source compiler framework. The SMT optimization for \xtalksched\ uses the Z3 SMT solver \cite{z3} version 4.8.4, using the Z3py APIs. We test our scheduler in comparison to two baselines \seriessched\ and \parsched, shown in Table \ref{tab:sched_variants}. \seriessched\ serializes all operations in the program. \parsched\ is the current state-of-the-art scheduler used in IBM Qiskit. 

\subsection{Benchmarks}
{\noindent \textbf{SWAP Circuits:}} 
We demonstrate the importance of crosstalk-adaptivity for communication orchestration in superconducting QC systems which have nearest-neighbor connectivity. In these architectures CNOTs are permitted only between adjacent qubits. To enable a CNOT between two far away qubits, compilers insert a sequence of SWAP operations which move the qubits into adjacent locations through exchanges. For example, in \ibmqpoughkeepsie, {\tt CNOT 0,13} can be implemented as {\tt SWAP 0,5; SWAP 5,10; SWAP 13,12; SWAP 12,11; CNOT 10,11;}, where both qubits meet-in-the-middle. Each SWAP operation is in turn composed of three CNOT gates\footnote{{\tt SWAP 0,1 := CNOT 0,1; CNOT 1,0; CNOT 0,1}}. Figure \ref{fig:swap_path_par} shows the operations executed for this sequence. 

We create meet-in-the-middle SWAP sequences between pairs of qubits in the device and schedule it using the three algorithms. When SWAP paths are executed on qubits which have no crosstalk e.g., on the path 0, 1, 2, 3 on \ibmqpoughkeepsie\ (see Figure \ref{fig:machine_figs}), \xtalksched\ and \parsched\ produce the same schedule. We avoid such SWAP paths in our evaluation and focus on $46$ circuits across the \numdevices\ devices which include at least one pair of high crosstalk CNOTs.

{\noindent \textbf{QAOA Circuits:}} We ran experiments on QAOA, a promising NISQ application, using the hardware efficient ansatz \cite{Moll_2018}. We used circuits with 4 qubits and 43 gates (9 two-qubit gates). We performed experiments on four crosstalk-prone regions in \ibmqpoughkeepsie.

{\noindent \textbf{Other benchmarks:}}
We also study our algorithm on the Hidden Shift benchmark \cite{hs} used in prior work \cite{triq, asplos}. We use Hidden Shift instances for sensitivity studies. Similar to SWAP circuits, we create instances of these circuits on subsets of qubits which are affected by crosstalk. To test scaling, we use instances of quantum supremacy circuits \cite{boxio}. 

\subsection{Metrics}
{\noindent \textbf{Crosstalk characterization:}} We count the number of SRB experiments and time required to perform characterization. We compare these metrics for a policy which performs SRB experiment all pairs of gates in the device, and with the three optimizations proposed in Section \ref{sec:xtalk_opt}.

{\noindent \textbf{Instruction scheduling:}} For SWAP circuits, we setup the circuit such that it creates a known answer, that can be measured (a Bell state) which can be measured using state tomography \cite{qiskit-ignis}. State tomography provides an error rate in the range $[0,1]$, with 1 meaning that the state is created perfectly. To execute state tomography we use 9216 trials (1024 per basis pair * 9 basis pairs) on the real system to obtain the error rate. 

For QAOA circuits, the output is obtained using 8192 trials. Since the output is a distribution of states, we used cross-entropy to measure the similarity of the output to the ideal theoretical distribution.
For Hidden Shift, we perform 8192 trials. The expected output is a single bit string, therefore, the error rate is measured as fraction of trials which did not yield the correct bit string. 

In all cases, readout error mitigation \cite{error-mitigation} is used to reduce the effect of imperfect hardware readout operations. 
\subsection{Setup}
Our compilation experiments use an Intel Core i7 processor (2.6GHz, 32GB RAM) with Python 3.7. We use \numdevices\ 20-qubit IBM systems for the quantum experiments. The device interface APIs in IBM Qiskit \cite{qiskit} were used to run characterization and application circuits. The daily machine calibration data is available through these APIs. The calibration data includes gate durations and independent error rates for all gates and coherence time ($T_1$ and $T_2$) for all qubits.

%% file: txt/results.tex
\section{Optimizing Application Error Rate} 
\begin{figure*}[t]
    \centering
    \subfloat[Error rates on \ibmqpoughkeepsie]
    {
    \includegraphics[scale=0.5]{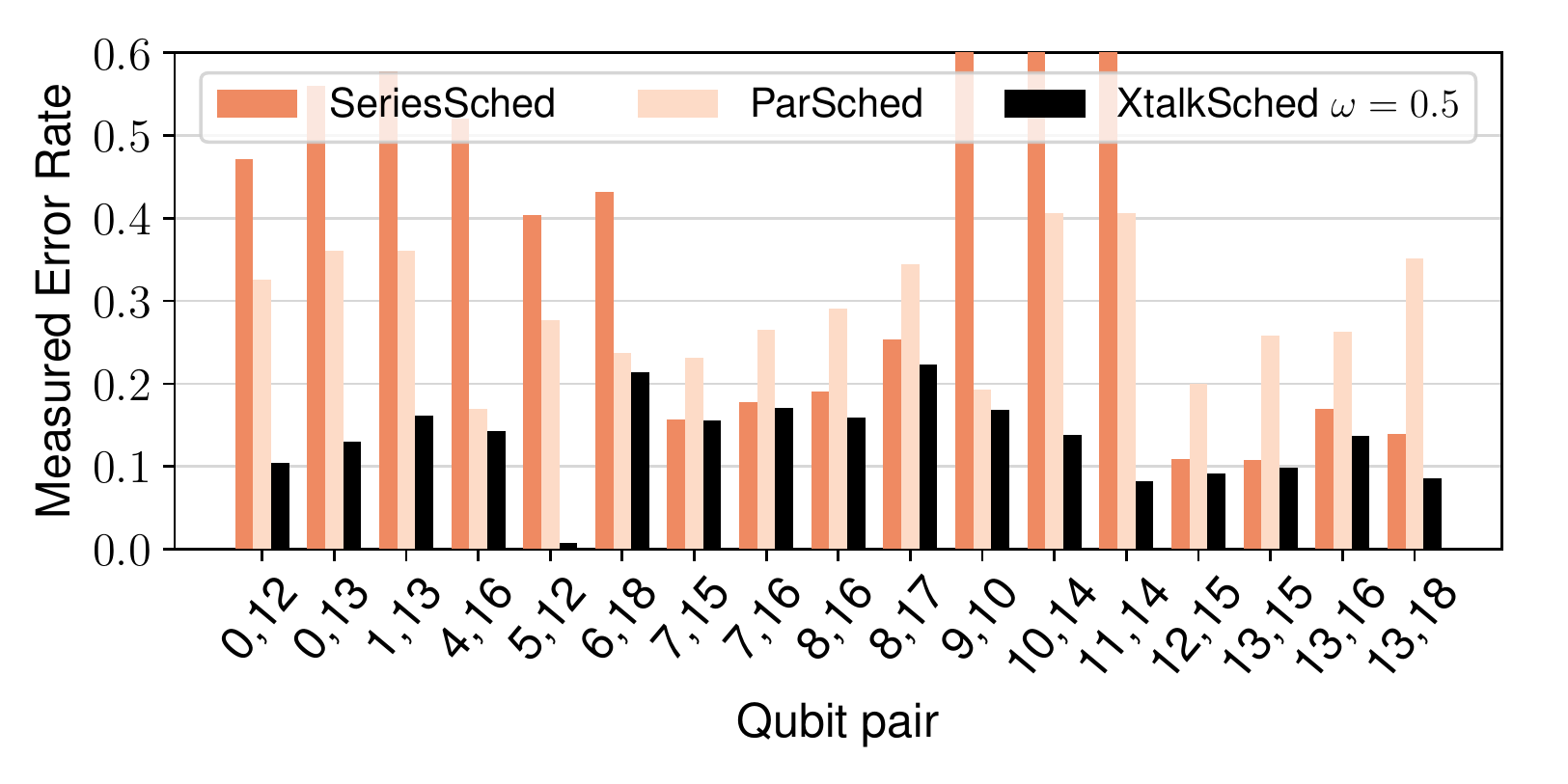}
    \label{fig:pougkeepsie_swap}
    }
    \subfloat[Error rates on \ibmqsystemone]
    {
    \includegraphics[scale=0.5]{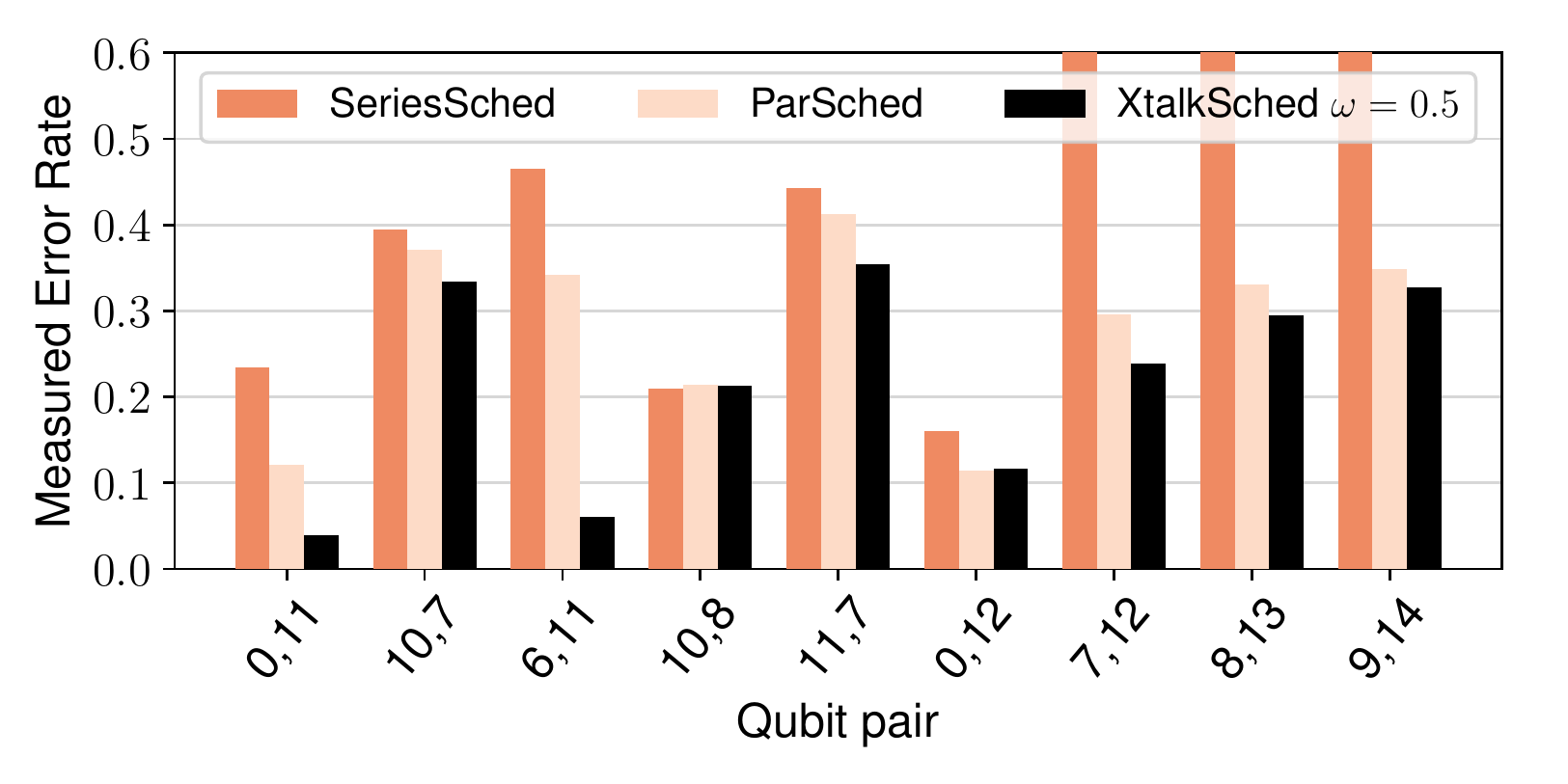}    
    \label{fig:systemone_swap}
    }
    
    \subfloat[Error rates on \ibmqboeblingen]
    {
    \includegraphics[scale=0.5]{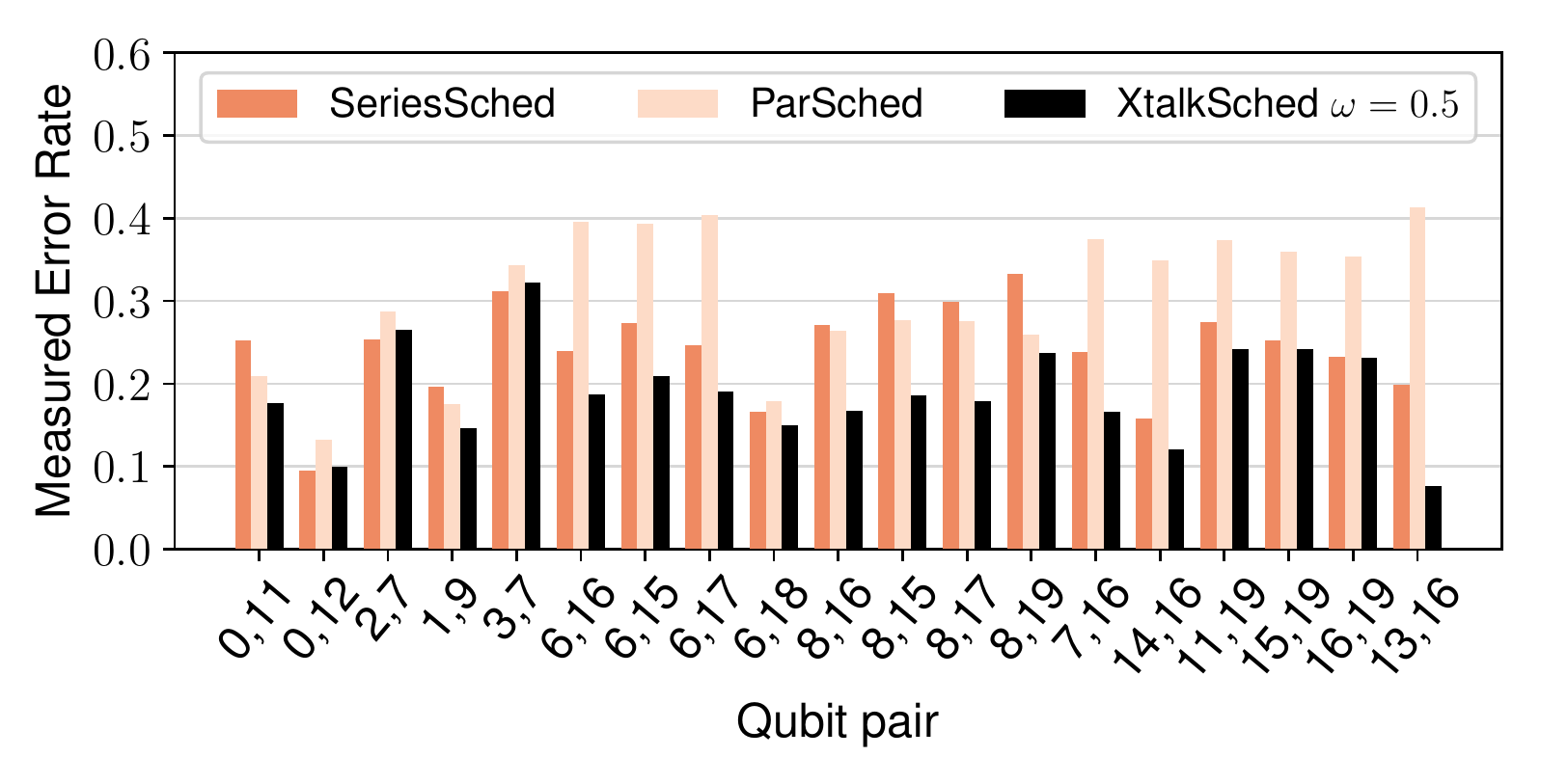}    
    \label{fig:boeblingen_swap}
    }
    \subfloat[Program durations on \ibmqpoughkeepsie]
    {
    \includegraphics[scale=0.5]{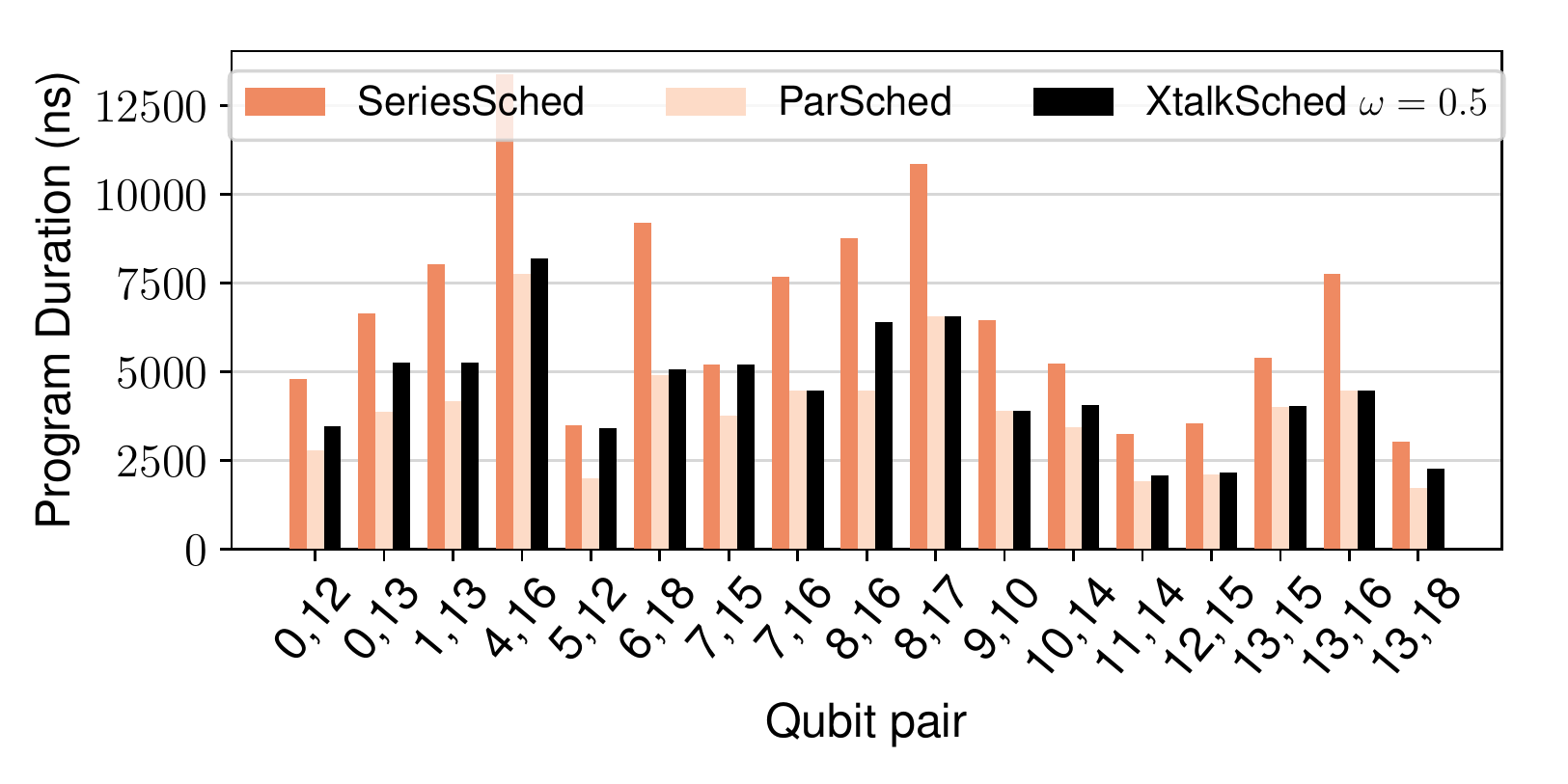}
    \label{fig:swap_duration_plot}
    }
    \caption{Figures (a), (b) and (c) show the measured error rates for SWAP circuits using the three schedulers on these systems. Lower error rate is better.
    \xtalksched\ obtains lower error across qubit pairs and across systems because it serializes high crosstalk operations while also optimizing the schedule to reduce the chances of decoherence. Figure (d) shows the program durations on \ibmqpoughkeepsie. Lower duration is better.
    With only a modest increase in execution time compared to \parsched, \xtalksched\ obtains large reductions in application error rate.}
    \label{fig:swap_plots}
\end{figure*}
\subsection{Comparisons to Baselines using SWAP Circuits}
{\noindent \textbf{Improvement in Error Rate:}}
Figure \ref{fig:swap_plots} compares the error rate for SWAP circuits scheduled with \xtalksched\ $\omega=0.5$, versus the \seriessched\ and \parsched\ schedulers, on the \numdevices\ systems. Although \seriessched\ naively serializes all instructions, in some cases it offers lower error than \parsched, because it avoids high crosstalk. \parsched\ outperforms \seriessched\ because it avoids decoherence by parallelizing operations. On all the tested qubit pairs, \xtalksched\ has significantly lower error rate than \seriessched\ and \parsched\ because it optimizes both crosstalk and decoherence. On \ibmqpoughkeepsie, \xtalksched\ obtains up to 4.9x reduction in error compared to \parsched, and up to 9.2x reduction compared to \seriessched. Across systems, the maximum improvement over \parsched\ is 5.6x, geomean 2x. 

{\noindent \textbf{Impact on Program Duration:}} We compare the durations of schedules produced by the three algorithms. Figure \ref{fig:swap_duration_plot} shows the program durations for SWAP circuits on \ibmqpoughkeepsie. Across different qubit pairs, \seriessched\ has the highest duration and \parsched\ has the lowest duration. \xtalksched\ produces executables which are only $1.16$x longer than \parsched\ on average, worst case $1.7$x. For NISQ applications, the most important figure of merit is the likelihood of correct execution, and not execution time. Nevertheless, \xtalksched\ needs to expend only a small increase in the execution time to mitigate crosstalk.

\begin{figure*}[t]
    \centering
    \subfloat[\seriessched]{
        \includegraphics[scale=0.14]{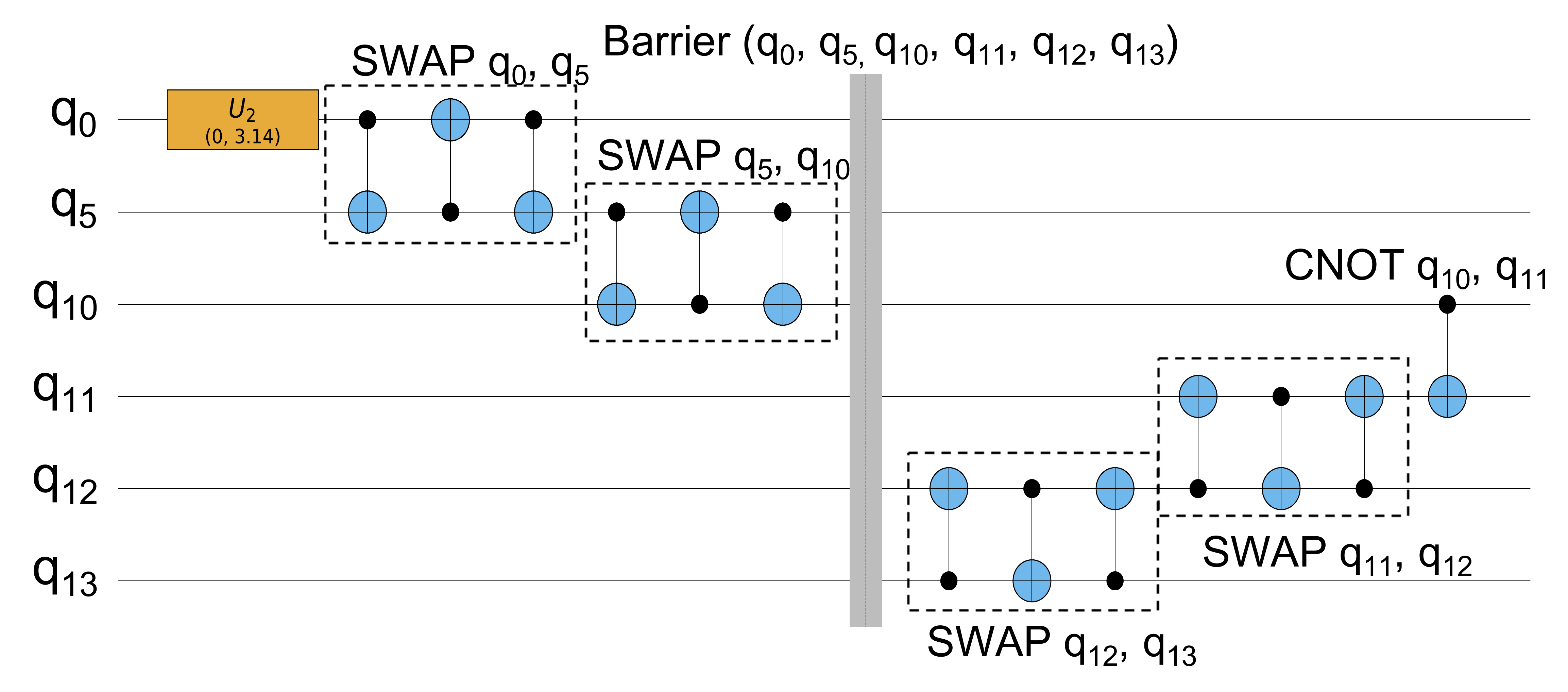}
    \label{fig:swap_path_ser}
        }
    \subfloat[\parsched]{
    \includegraphics[scale=0.14]{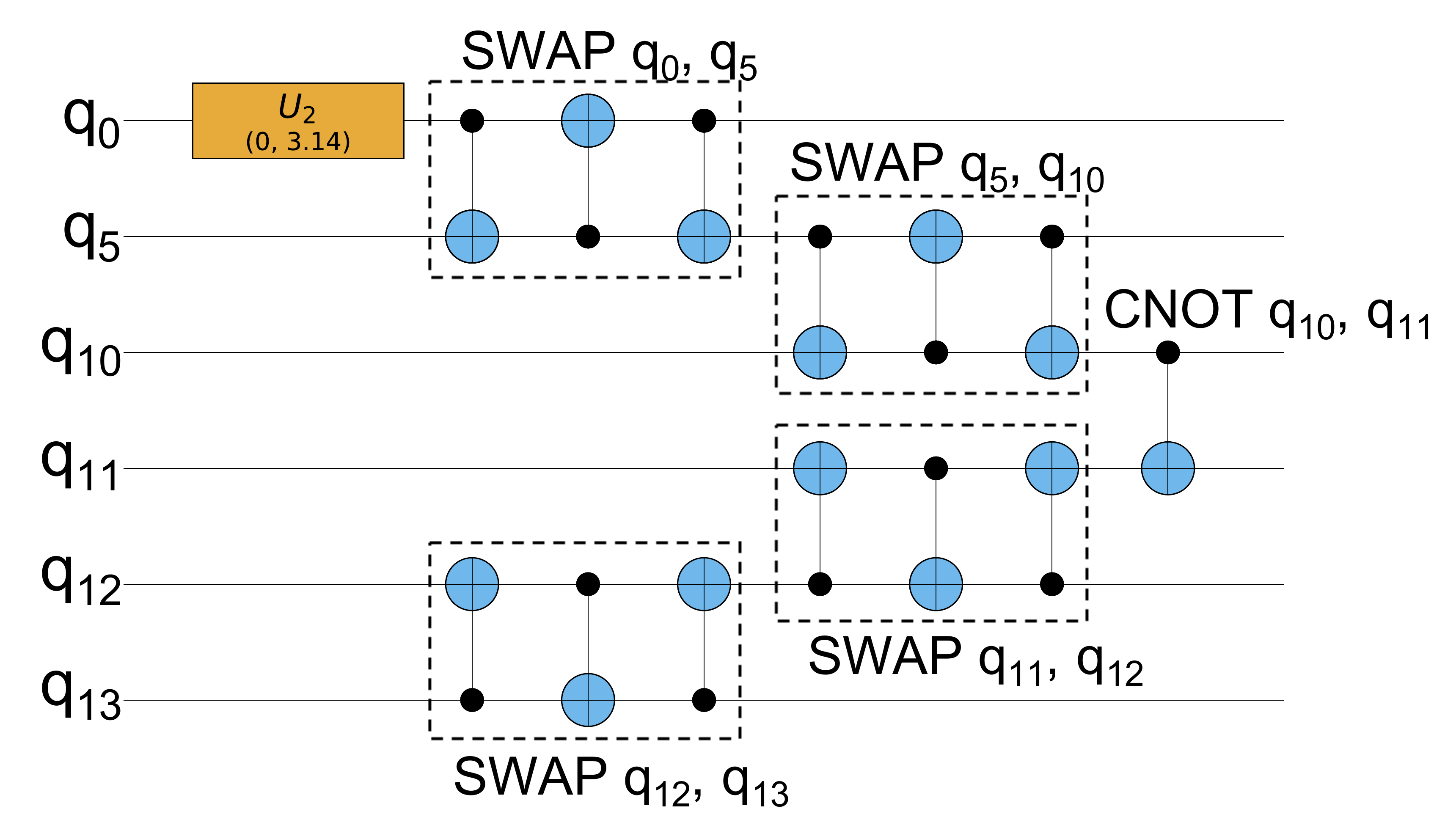}
    \label{fig:swap_path_par}
    }
    \subfloat[\xtalksched]{
        \includegraphics[scale=0.14]{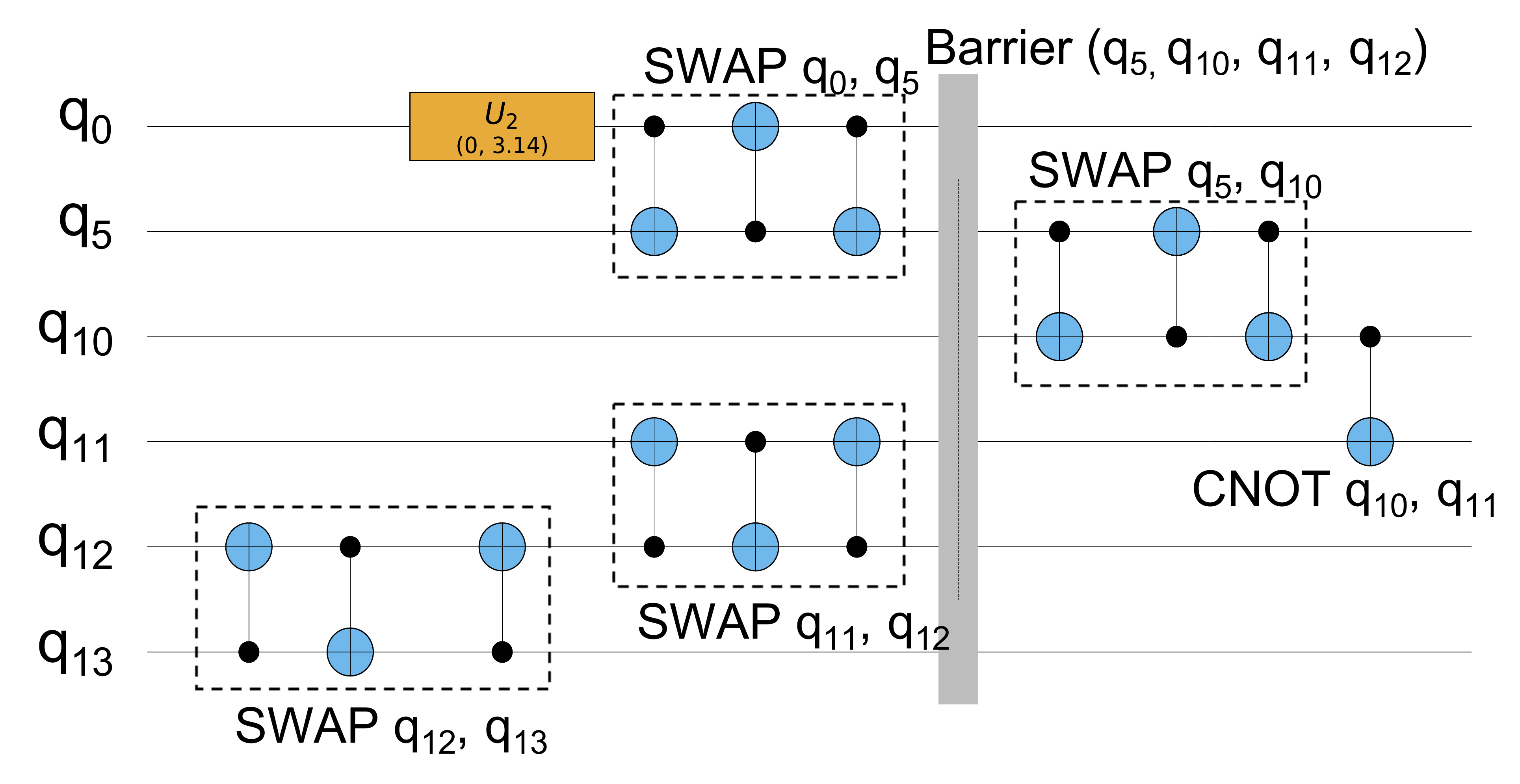}
    \label{fig:swap_path_xtalk}
    }
    \caption{Schedules produced by the 3 algorithms for the SWAP path between qubit 0 and qubit 13 in \ibmqpoughkeepsie. All schedules have a U2 operation on qubit 0, required for creating a known final answer. \seriessched\ serializes all the 4 SWAP operations. SWAP 0,5 and SWAP 5, 10 are serialized because of the data dependency; they are serialized with the remaining swaps because of the barrier. Because of full serialization, \seriessched\ incurs high decoherence error. \parsched\ executes maximum gates in parallel in each timestep and incurs high crosstalk errors for the three simultaneous CNOT 5, 10 and CNOT 11, 12 gates (second set of parallel SWAPs). \xtalksched\ optimizes crosstalk errors by serializing operations and mitigates the chances of decoherence by ordering gates to reduce the durations on qubits which have low coherence times.}
    \label{fig:swap_path_illustration}
\end{figure*}

{\noindent \textbf{Example Case:}}
Figure \ref{fig:swap_path_illustration} shows the schedules for the swap path between qubit 0 and 13 on \ibmqpoughkeepsie. \seriessched\ schedules all 4 SWAPs in series and avoids crosstalk errors. But, it has high schedule length and therefore, high decoherence error. \parsched\ schedules the two pairs of logically independent SWAPs in parallel which reduces the execution time and decoherence errors. But, it incurs high crosstalk errors for the SWAP operation on qubits 5, 10 and the SWAP on 11, 12. \xtalksched\ obtains the best of both cases. It parallelizes the far away SWAPs which don't have crosstalk, and serializes the nearby SWAPs. This allows it to avoid the crosstalk noise, which compensates for a small increase in decoherence and improves the overall error rate. 

For serializing the two SWAPs \xtalksched\ chooses the best ordering of operations i.e., when two gates $g_i$ and $g_j$ need to be serialized, it decides whether $g_i$ should be placed before or after $g_j$. For this system, qubit $10$ has very low coherence time (less than $6$us, which is nearly 10X lower than the average coherence on this system). On the IBM systems, decoherence effects on a qubit start only after the first gate is applied. If \xtalksched\ performs SWAP 5,10 first, followed by SWAP 11,12, the state of qubit 10 would decohere. Instead, since we model qubit lifetime to start at the first gate (constraint \ref{lifetime_constraint}), \xtalksched\ computes an optimal ordering where SWAP 5,10 gets placed with minimum lifetime, after SWAP 11,12.

{\noindent \textbf{Optimality:}} For qubits affected by crosstalk on \ibmqpoughkeepsie, Figure \ref{fig:swap_optimality} compares \xtalksched\ swap error rates to the ideal crosstalk-free error rates. To obtain the ideal error rates, we averaged swap error rates on crosstalk-free swap paths in \ibmqpoughkeepsie, selecting the lowest error schedule for each path. Figure \ref{fig:swap_optimality} shows that XtalkSched error rates are close to the ideal and within geomean $1\% \pm 16\%$ (1 standard deviation) of average error rate of crosstalk-free swap paths of the same length. 

{\em Given fundamental connectivity restrictions on superconducting QC systems, SWAP-based communication is important for all programs run on these systems, especially as devices and programs scale up. XtalkSched's near-optimal crosstalk mitigation and improved error rate is therefore very relevant for reliable execution on current and near-term NISQ systems.}
\begin{figure}
\begin{tabular}{|r|r|r|r|}
\hline
\begin{tabular}[c]{@{}c@{}}Qubit\\ Pair\end{tabular} & \begin{tabular}[c]{@{}c@{}}XtalkSched\\ Error Rate\end{tabular} & \begin{tabular}[c]{@{}c@{}}Ideal Error Rate \\ (Crosstalk Free)\end{tabular} & \begin{tabular}[c]{@{}c@{}}Path\\ Length\end{tabular} \\ \hline
5, 12                                                & 0.007                                                           & \multirow{4}{*}{$0.100 \pm 0.068$}                                          & \multirow{4}{*}{3}                                    \\ \cline{1-2}
11, 14                                               & 0.081                                                           &                                                                             &                                                       \\ \cline{1-2}
12, 15                                               & 0.091                                                           &                                                                             &                                                       \\ \cline{1-2}
13, 18                                               & 0.085                                                           &                                                                             &                                                       \\ \hline
0, 12                                                & 0.104                                                           & \multirow{4}{*}{$0.133 \pm 0.067$}                                          & \multirow{4}{*}{4}                                    \\ \cline{1-2}
7, 15                                                & 0.154                                                           &                                                                             &                                                       \\ \cline{1-2}
10, 14                                               & 0.138                                                           &                                                                             &                                                       \\ \cline{1-2}
13, 15                                               & 0.098                                                           &                                                                             &                                                       \\ \hline
0, 13                                                & 0.129                                                           & \multirow{4}{*}{$0.162 \pm 0.057$}                                          & \multirow{4}{*}{5}                                    \\ \cline{1-2}
7, 16                                                & 0.171                                                           &                                                                             &                                                       \\ \cline{1-2}
9, 10                                                & 0.167                                                           &                                                                             &                                                       \\ \cline{1-2}
13, 16                                               & 0.137                                                           &                                                                             &                                                       \\ \hline
1, 13                                                & 0.161                                                           & \multirow{3}{*}{$0.179 \pm 0.067$}                                          & \multirow{3}{*}{6}                                    \\ \cline{1-2}
6, 18                                                & 0.213                                                           &                                                                             &                                                       \\ \cline{1-2}
8, 16                                                & 0.158                                                           &                                                                             &                                                       \\ \hline
8, 17                                                & 0.222                                                           & $0.230 \pm 0.04$                                                            & 7                                                     \\ \hline
4, 16                                                & 0.143                                                           & $0.251 \pm 0.04$                                                            & 8                                                     \\ \hline
\end{tabular}
\caption{Comparison of XtalkSched error rates to ideal swap error rates (lower is better). Experiments on \ibmqpoughkeepsie. Ideal error rates are measured by averaging swap error rates on crosstalk-free paths, selecting the lowest error schedules. XtalkSched error rates are very close to the ideal error rates, indicating near-optimal crosstalk mitigation.}
\label{fig:swap_optimality}
\end{figure}
\begin{figure}
    \centering
    \includegraphics[scale=0.5]{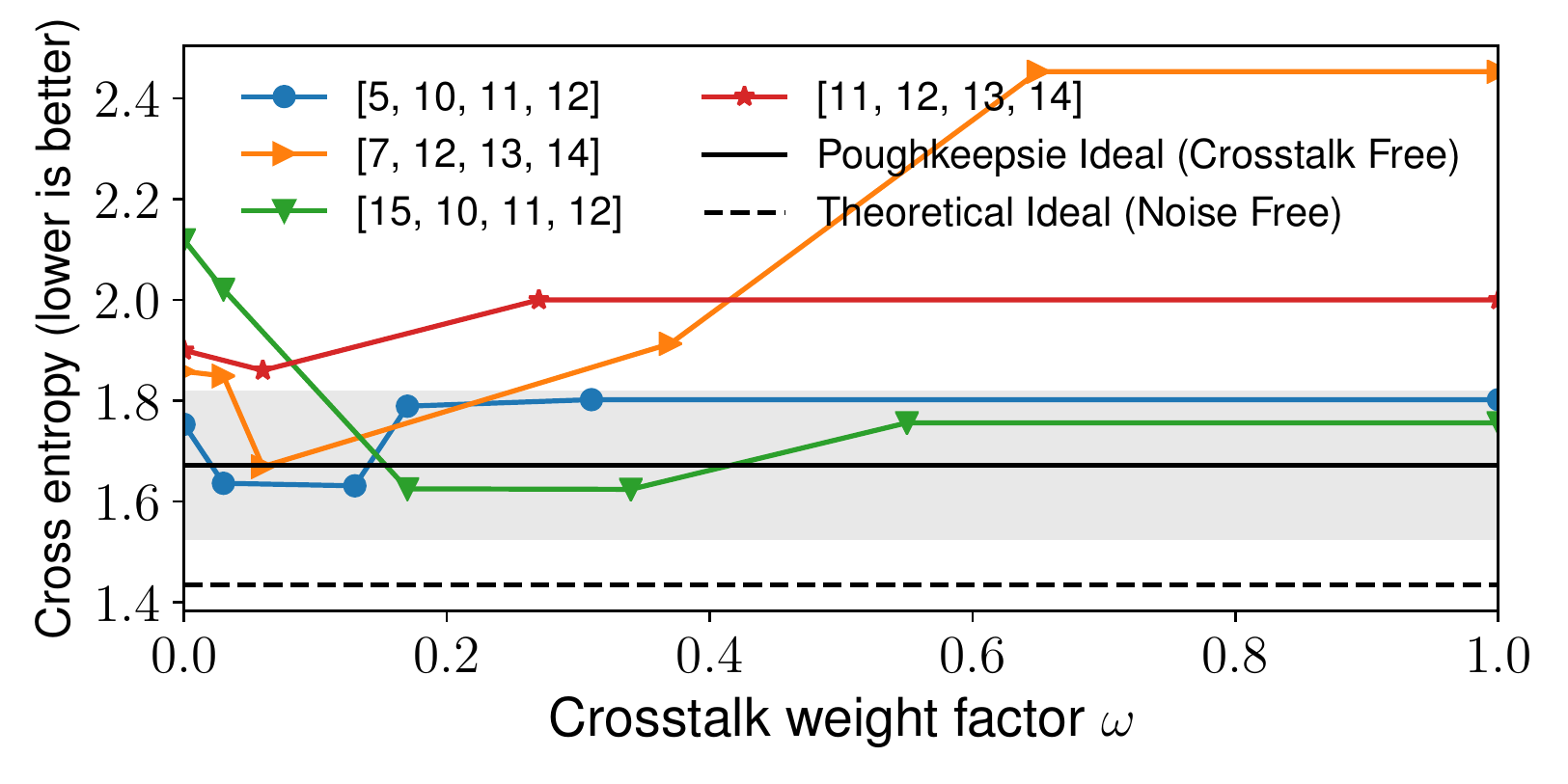}
    \caption{Measured cross entropy for QAOA circuits using \xtalksched, $\omega \in [0,1]$. Lower cross entropy is better. The dotted line indicates the perfect noise-free cross entropy. The solid black line (mean) and the grey band (1 standard deviation) indicate the cross entropy achievable on crosstalk-free regions of the device. XtalkSched reduces cross-entropy loss by geomean 1.8x (up to 3.6x) compared to ParSched ($\omega=0$) and geomean 2x (up to 4.3x) compared to SeriesSched ($\omega=1$). Cross entropy offered by XtalkSched is very close to or within the ideal range.}
    \label{fig:qaoa_res}
\end{figure}
\begin{figure}
    \centering
    \subfloat[Hidden Shift with no redundant CNOTs: Less susceptible to crosstalk]{
    \includegraphics[scale=0.5]{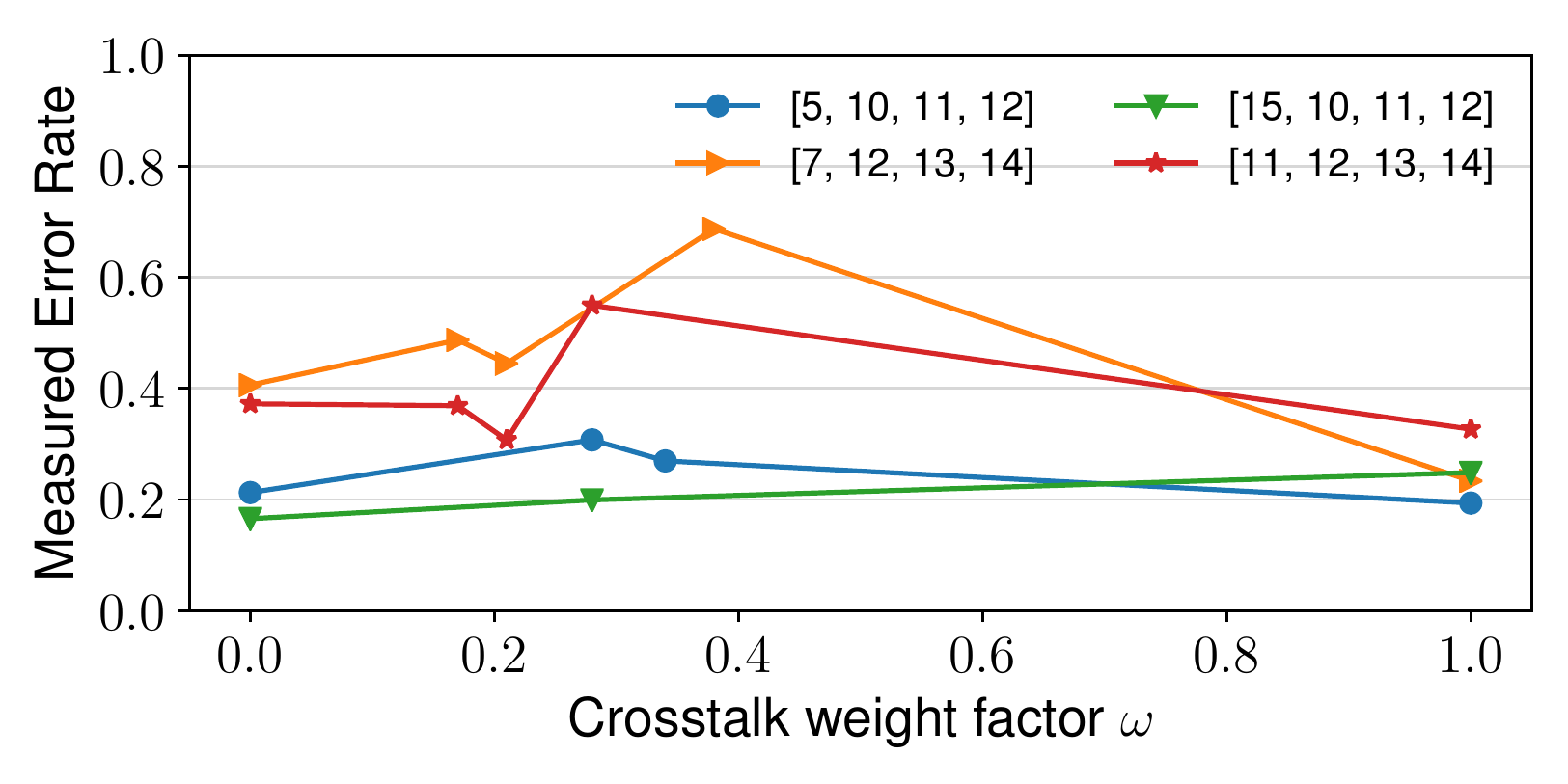}
    \label{fig:hs_results_no_red}
    }
    
    \subfloat[Hidden shift with redundant CNOTs: More susceptible to crosstalk]{
        \includegraphics[scale=0.5]{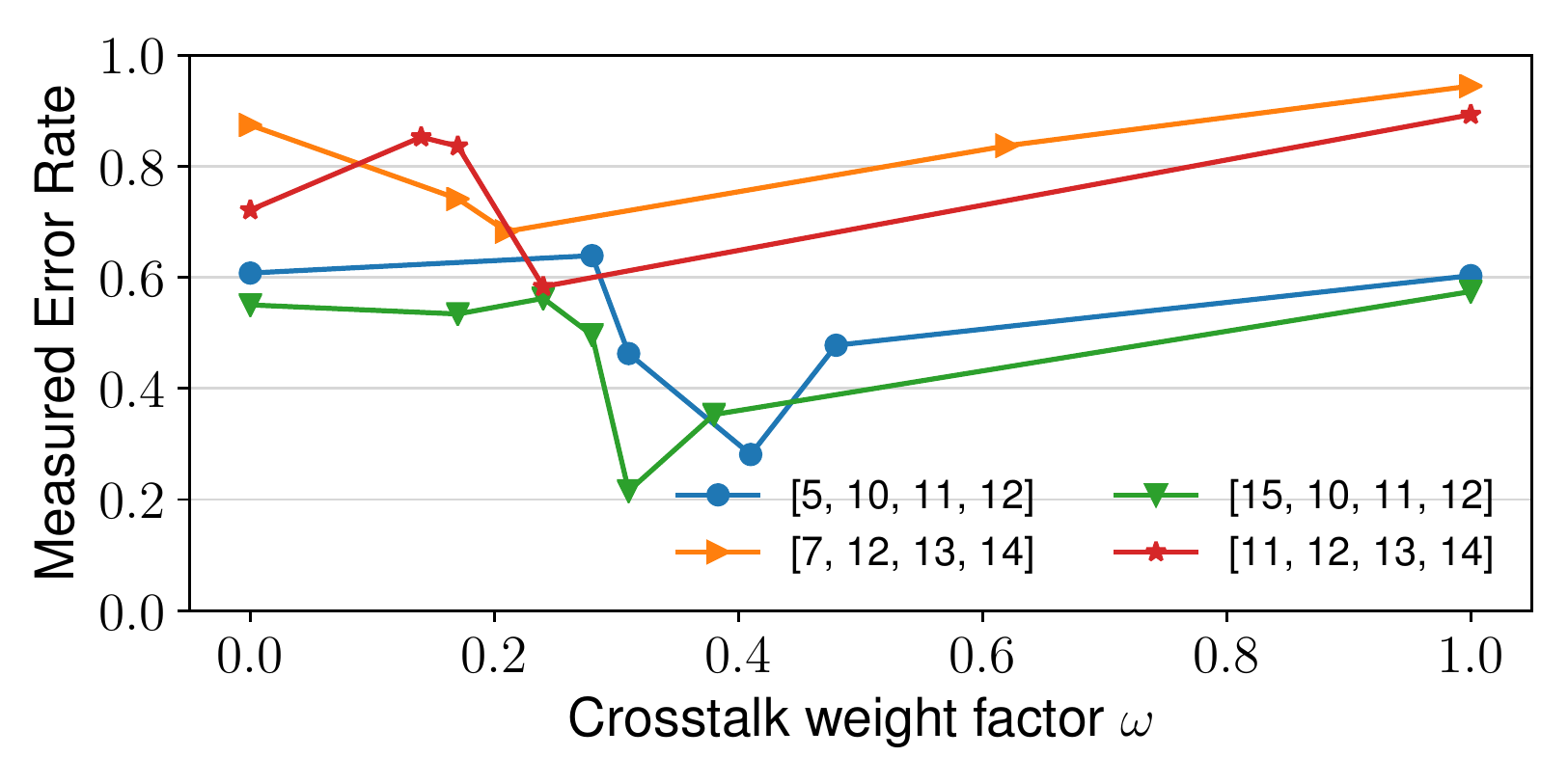}
        \label{fig:hs_results_red}
    }
    \caption{Sensitivity of \xtalksched\ to the choice of $\omega$. Figure (a) shows that when the application is less susceptible to crosstalk noise, the choice of weight factor matters a lot. Only when $\omega=1$, i.e., consider only crosstalk and ignore decoherence, we beat the baseline with $\omega=0$ (maximize parallelism). Figure (b) shows that when the application is more susceptible to crosstalk noise, any $\omega \in [0.2, 0.5]$ beats the baseline with $\omega=0$.}
    \label{fig:hs_results}
\end{figure}


\subsection{Evaluation on QAOA Circuits}
Figure \ref{fig:qaoa_res} shows the cross entropy for QAOA circuits on \ibmqpoughkeepsie\ using \xtalksched\ with $\omega \in [0,1]$. Cross entropy measures how close the output distribution is to the ideal distribution obtained from a noise-free simulation with Qiskit Aer simulator. 
With $\omega=1$ \xtalksched\ considers only crosstalk noise and ignores decoherence. Hence it serializes all instructions similar to \seriessched. With $\omega=0$, only decoherence is considered, and \xtalksched\ is equivalent to \parsched. When $\omega$ is varied from 0.03 to 0.2, \xtalksched\ outperforms both the baselines and significantly reduces the cross entropy. 
XtalkSched reduces the loss in cross-entropy (with respect to the ideal) by geomean 1.8x (up to 3.6x) compared to ParSched and geomean 2x (up to 4.3x) compared to SeriesSched. Further, we performed experiments on crosstalk-free regions of the hardware to measure the average cross entropy achievable on the device. Owing to variability in gate errors across the device, this value has mean 1.67 and standard deviation 0.15 and is indicated by the grey band in Figure \ref{fig:qaoa_res}. XtalkSched offers cross entropy within this ideal band and hence the crosstalk mitigation is near-optimal.

\subsection{Sensitivity of Weight Factor to Application Characteristics}
We test the relationship between the susceptibility of an application to crosstalk noise and the choice of $\omega$ using the Hidden Shift benchmark.
 This benchmark has 2 layers of parallel CNOTs, with each layer containing 2 CNOTs. Because of variable gate durations, the CNOTs in the benchmark may not fully overlap or have high interference. We make this benchmark susceptible to crosstalk by replacing each CNOT gate by three CNOT gates. The first two CNOT gates act as an identity gate, but they have crosstalk noise with other parallel gates. 

We performed experiments on four instances of HS varying the weights. Figure \ref{fig:hs_results} shows the results. For the original benchmark, the results are highly sensitive to the crosstalk weight factor. Only \xtalksched\ with $\omega=1$ (only consider crosstalk, ignore decoherence) obtains improvements over \xtalksched\ with $\omega=0$ (parallelize all operations). In contrast, when we introduce redundant CNOTs, any weight from $0.2$ to $0.5$ improves over $\omega=0$. The best case improvements are high as 3x. Hence, for applications which are very susceptible to crosstalk noise, it is easy to obtain improvements from our scheduler even without very careful tuning. 

\subsection{Scalability Study}
We tested the scalability of our scheduler on instances of quantum supremacy benchmarks \cite{boxio}. These programs are random circuits which are hard to simulate classically. We tested instances with 6-18 qubits, with 100 to 1000 gates (depth 40). Our algorithm's scaling behavior depends on the number of gates rather than the number of qubits, since the constraints are primarily on the gate schedule. In instances with 18 qubits and 500 gates, the compilation time is less than 2 minutes. Even with 1000 gates, the compile times are under 15 minutes. These execution times can be easily improved with known optimizations for SMT compilers \cite{asplos, triq}. This evaluation gives us confidence that our methods will be practical even on large NISQ-era workloads. 

\section{Fast Crosstalk Characterization}
\begin{figure}
    \centering
    \includegraphics[scale=0.6]{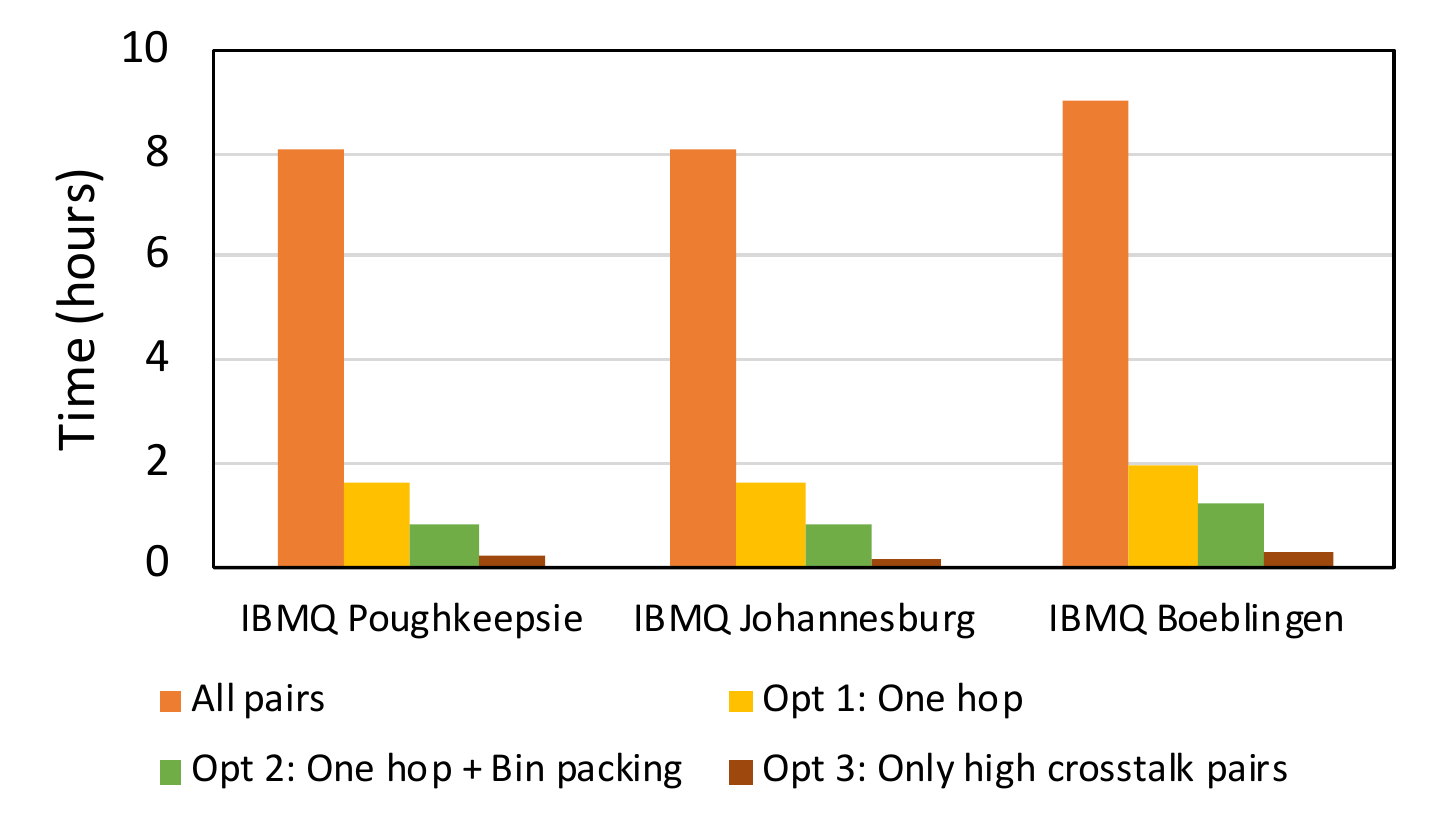}
    \caption{Crosstalk characterization time for 3 systems. Lower time is better. With our optimizations, the characterization overhead is significantly reduced, allowing frequent characterization to support the compiler optimization.}
    \label{fig:fastcrosstalkexpt}
\end{figure}

Figure \ref{fig:fastcrosstalkexpt} shows the time required for crosstalk characterization using the policies discussed in Section \ref{sec:xtalk_opt}. In the baseline policy, when crosstalk among all pairs of hardware CNOTs is characterized, the number of experiments required is as high as 246. Each such experiment requires thousands of random trials and the overall execution time for characterizing crosstalk once becomes well over 8 hours per system. Thus, a third of the system's overall lifetime will be spent in measuring crosstalk. With our first optimization, we can restrict measurements to gate pairs separated by 1 hop and reduce the overhead by 5x. By parallelizing these experiments using bin packing, our second optimization provides a further 2x reduction. Finally, by observing that the conditional error rates of high crosstalk gates are sufficient for compilation, we obtain a further 4-7x reduction across systems. Overall, our optimizations reduce the number of experiments by 35-73x across the 3 systems over the baseline policy, allowing us to frequently characterize crosstalk and provide accurate inputs to the scheduler.

%% file: txt/conclusions.tex


\section{Conclusions}
We develop and demonstrate an approach for software mitigation of crosstalk noise in NISQ systems. To this end, we developed a fast and accurate crosstalk characterization methods, and an instruction scheduler which yields up to 5.6x better error than state-of-the-art compilers on 3 20-qubit IBM quantum systems and several application benchmarks. Our work shows that crosstalk mitigation in software is possible, and can greatly increase the reliability of noisy quantum computers.

Our scheduler is available open source at \url{https://github.com/Qiskit/qiskit-terra/blob/master/qiskit/transpiler/passes/crosstalk_adaptive_schedule.py}.




